\documentclass[fleqn,usenatbib]{mnras}

\usepackage{newtxtext,newtxmath}
\usepackage[T1]{fontenc}
\usepackage{lipsum}
\DeclareRobustCommand{\VAN}[3]{#2}
\let\VANthebibliography\thebibliography
\def\thebibliography{\DeclareRobustCommand{\VAN}[3]{##3}\VANthebibliography}

\usepackage{graphicx}	% Including figure files
\usepackage{amsmath}	% Advanced maths commands
\usepackage{multirow}
\usepackage{array}
\usepackage{makecell}
\usepackage{blindtext}
\usepackage{siunitx}
\usepackage{booktabs}
\usepackage{subcaption}
\title[Constraining the phase shift in DESI BAOs]{\vspace{-1.5em}Constraining the phase shift of relativistic species in DESI BAOs\vspace{-0.75em}}

\author[A. M. Whitford et al.]{Abbé M. Whitford,$^{1}$\thanks{abbe.whitford@gmail.com}
Hugo Rivera-Morales,$^{2}$
Cullan Howlett,$^{1}$
Mariana Vargas-Magaña,$^{2}$
\newauthor
Sébastien Fromenteau,$^{3}$
Tamara M. Davis,$^{1}$
Alejandro Pérez-Fernández,$^{4}$
Arnaud de Mattia,$^{5}$
Steven Ahlen,$^{6}$
\newauthor 
Davide Bianchi,$^{7}$
David Brooks,$^{8}$
Etienne Burtin,$^{5}$
Todd Claybaugh,$^{9}$
Axel de la Macorra,$^{2}$
Peter Doel,$^{8}$
\newauthor 
Simone Ferraro,$^{10,9}$
Jaime E. Forero-Romero,$^{11,12}$
Enrique Gaztañaga,$^{15,13,14}$
Satya Gontcho A Gontcho,$^{9}$
\newauthor 
Gaston Gutierrez,$^{16}$
Stephanie Juneau,$^{17}$
Robert Kehoe,$^{18}$
David Kirkby,$^{19}$
Theodore Kisner,$^{9}$
\newauthor 
Sergey Koposov,$^{20,21}$
Martin Landriau,$^{9}$
Laurent Le Guillou,$^{22}$
Aaron Meisner,$^{17}$
Ramon Miquel,$^{24,23}$
\newauthor 
Francisco Prada,$^{25}$
Ignasi Pérez-Ràfols,$^{26}$
Graziano Rossi,$^{27}$
Eusebio Sanchez,$^{28}$
Michael Schubnell,$^{30,29}$
\newauthor 
David Sprayberry,$^{17}$
Gregory Tarlé,$^{30}$
Benjamin Alan Weaver,$^{17}$
Pauline Zarrouk,$^{22}$
Hu Zou$^{31}$
\\
$^{1}$School of Mathematics and Physics, University of Queensland, Brisbane, QLD 4072, Australia\\
$^{2}$Universidad Nacional Aut\'{o}noma de M\'{e}xico. Instituto de F\'{\i}sica.  A.P. 20-364, 01000. Cd. de M\'{e}xico,  M\'{e}xico\\
$^{3}$Universidad Nacional Autónoma de México. Instituto de Ciencias Físicas, 62210, Cuernavaca, Morelos, México\\
$^{4}$Max Planck Institute for Extraterrestrial Physics, Gie\ss enbachstra\ss e 1, 85748 Garching, Germany\\
$^{5}$IRFU, CEA, Universit\'{e} Paris-Saclay, F-91191 Gif-sur-Yvette, France\\
$^{6}$Physics Dept., Boston University, 590 Commonwealth Avenue, Boston, MA 02215, USA\\
$^{7}$Dipartimento di Fisica ``Aldo Pontremoli'', Universit\`a degli Studi di Milano, Via Celoria 16, I-20133 Milano, Italy\\
$^{8}$Department of Physics \& Astronomy, University College London, Gower Street, London, WC1E 6BT, UK\\
$^{9}$Lawrence Berkeley National Laboratory, 1 Cyclotron Road, Berkeley, CA 94720, USA\\
$^{10}$University of California, Berkeley, 110 Sproul Hall \#5800 Berkeley, CA 94720, USA\\
$^{11}$Departamento de F\'isica, Universidad de los Andes, Cra. 1 No. 18A-10, Edificio Ip, CP 111711, Bogot\'a, Colombia\\
$^{12}$Observatorio Astron\'omico, Universidad de los Andes, Cra. 1 No. 18A-10, Edificio H, CP 111711 Bogot\'a, Colombia\\
$^{13}$Institut d'Estudis Espacials de Catalunya (IEEC), c/ Esteve Terradas 1, Edifici RDIT, Campus PMT-UPC, 08860 Castelldefels, Spain\\
$^{14}$Institute of Cosmology and Gravitation, University of Portsmouth, Dennis Sciama Building, Portsmouth, PO1 3FX, UK\\
$^{15}$Institute of Space Sciences, ICE-CSIC, Campus UAB, Carrer de Can Magrans s/n, 08913 Bellaterra, Barcelona, Spain\\
$^{16}$Fermi National Accelerator Laboratory, PO Box 500, Batavia, IL 60510, USA\\
$^{17}$NSF NOIRLab, 950 N. Cherry Ave., Tucson, AZ 85719, USA\\
$^{18}$Department of Physics, Southern Methodist University, 3215 Daniel Avenue, Dallas, TX 75275, USA\\
$^{19}$Department of Physics and Astronomy, University of California, Irvine, 92697, USA\\
$^{20}$Institute for Astronomy, University of Edinburgh, Royal Observatory, Blackford Hill, Edinburgh EH9 3HJ, UK\\
$^{21}$Institute of Astronomy, University of Cambridge, Madingley Road, Cambridge CB3 0HA, UK\\
$^{22}$Sorbonne Universit\'{e}, CNRS/IN2P3, Laboratoire de Physique Nucl\'{e}aire et de Hautes Energies (LPNHE), FR-75005 Paris, France\\
$^{23}$Instituci\'{o} Catalana de Recerca i Estudis Avan\c{c}ats, Passeig de Llu\'{\i}s Companys, 23, 08010 Barcelona, Spain\\
$^{24}$Institut de F\'{i}sica d’Altes Energies (IFAE), The Barcelona Institute of Science and Technology, Edifici Cn, Campus UAB, 08193, Bellaterra (Barcelona), Spain\\
$^{25}$Instituto de Astrof\'{i}sica de Andaluc\'{i}a (CSIC), Glorieta de la Astronom\'{i}a, s/n, E-18008 Granada, Spain\\
$^{26}$Departament de F\'isica, EEBE, Universitat Polit\`ecnica de Catalunya, c/Eduard Maristany 10, 08930 Barcelona, Spain\\
$^{27}$Department of Physics and Astronomy, Sejong University, 209 Neungdong-ro, Gwangjin-gu, Seoul 05006, Republic of Korea\\
$^{28}$CIEMAT, Avenida Complutense 40, E-28040 Madrid, Spain\\
$^{29}$Department of Physics, University of Michigan, 450 Church Street, Ann Arbor, MI 48109, USA\\
$^{30}$University of Michigan, 500 S. State Street, Ann Arbor, MI 48109, USA\\
$^{31}$National Astronomical Observatories, Chinese Academy of Sciences, A20 Datun Rd., Chaoyang District, Beijing, 100012, P.R. China\\
\vspace{-3em}}

\date{Accepted XXX. Received YYY; in original form ZZZ\vspace{-1.75em}}

\pubyear{\the\year{}}

\begin{document}

\label{firstpage}
\pagerange{\pageref{firstpage}--\pageref{lastpage}}
\maketitle

\begin{abstract}
In the early Universe, neutrinos decouple quickly from the primordial plasma and propagate without further interactions. The impact of free-streaming neutrinos is to create a temporal shift in the gravitational potential that impacts the acoustic waves known as baryon acoustic oscillations (BAOs), resulting in a non-linear spatial shift in the Fourier-space BAO signal. In this work, we make use of and extend upon an existing methodology to measure the phase shift amplitude $\beta_{\phi}$ and apply it to the DESI Data Release 1 (DR1) BAOs with an anisotropic BAO fitting pipeline. We validate the fitting methodology by testing the pipeline with two publicly available fitting codes 
%used by the DESI Collaboration, \textit{desilike} and \textit{Barry}. We 
applied to highly precise cubic box simulations and realistic simulations representative of the DESI DR1 data. We find further study towards the methods used in fitting the BAO signal 
%broadband power spectrum 
will be necessary to ensure accurate constraints on $\beta_{\phi}$ in future DESI data releases. Using DESI DR1, we present individual measurements of the anisotropic BAO distortion parameters and the $\beta_{\phi}$ for the different tracers, and additionally a combined fit to $\beta_{\phi}$ resulting in $\beta_{\phi} = 2.7 \pm 1.7$. After including a prior on the distortion parameters from constraints using \textit{Planck} we find $\beta_{\phi} = 2.7^{+0.60}_{-0.67} $ suggesting $\beta_{\phi} > 0$ at 4.3$\sigma$ significance. This result may hint at a phase shift that is not purely sourced from the standard model expectation for $N_{\rm{eff}}$ or could be a upwards statistical fluctuation in the measured $\beta_{\phi}$; this result relaxes in models with additional freedom beyond $\Lambda$CDM. 
%which cannot allow $\beta_{\phi} \gtrsim 2.44$. 
\end{abstract}



%%%%%%%%%%%%%%%%%%%%%%%%%%%%%%%%%%%%%%%%%%%%%%%%%%

%%%%%%%%%%%%%%%%% BODY OF PAPER %%%%%%%%%%%%%%%%%%

\section{Introduction}

The present-day distribution of matter perturbations in the Universe has grown from the over and under-densities that froze in place when sound waves in the early Universe ceased to propagate. These sound waves are called the BAOs, and could propagate in the hot and dense plasma at early times due to the coupling between baryons and photons. At the epoch of recombination at $z \sim 1090$, photons were no longer energetic enough to ionize atoms, and photons began decouple from the baryons and stream freely; these photons are measured today as the cosmic microwave background \citep[CMB, ][]{penzias1979measurement}. The BAOs ceased to propagate completely at the baryon drag epoch at which baryons and photons completely decoupled.

In the present day, galaxies throughout the Universe trace the underlying matter distribution; galaxy surveys such as the Dark Energy Spectroscopic Instrument \citep[DESI,][]{aghamousa2016desi, aghamousa2016desi2, abareshi2022overview, adame2024desi} aim to probe the large-scale distribution of galaxies to measure the imprint of the BAOs and test models of cosmology. This has already been probed by various surveys \citep{cole20052df, eisenstein2005detection, gil2016clustering, abbott2019dark} and most recently in DESI data \citep{moon2023first, adame2024desi3, adame2024desi} who have measured the `peak' in the two-point auto-correlation function of galaxies or the `wiggles' that can be seen due to this peak in Fourier space (the Fourier-space transform is referred to as the power spectrum). The peak corresponds to the BAO scale $r_s$; the longest distance sound waves could travel up until the baryon drag epoch and also represents a separation scale at which there is an excess probability of finding pairs of galaxies. 

Historically, the BAO peak is used to to test cosmological models in a partially model-independent way; by measuring a parameter $\alpha$ that represents a ratio of distance scales in an fiducial cosmological model (used to convert the galaxy catalogue angular coordinates and redshifts into Cartesian coordinates) to the same distance scales in the true cosmology. If the fiducial cosmology matches the truth, the parameter $\alpha$ is unity; if not, deviations from unity occur as the physical BAO scale and the BAO scale of the fiducial cosmology do not match. The scaling of the parameter $\alpha$ physically represents an isotropic scaling or distortion of the distance to the BAO peak \citep{bernal2020robustness}.

\begin{figure}
    \centering \includegraphics[width=0.48\textwidth]{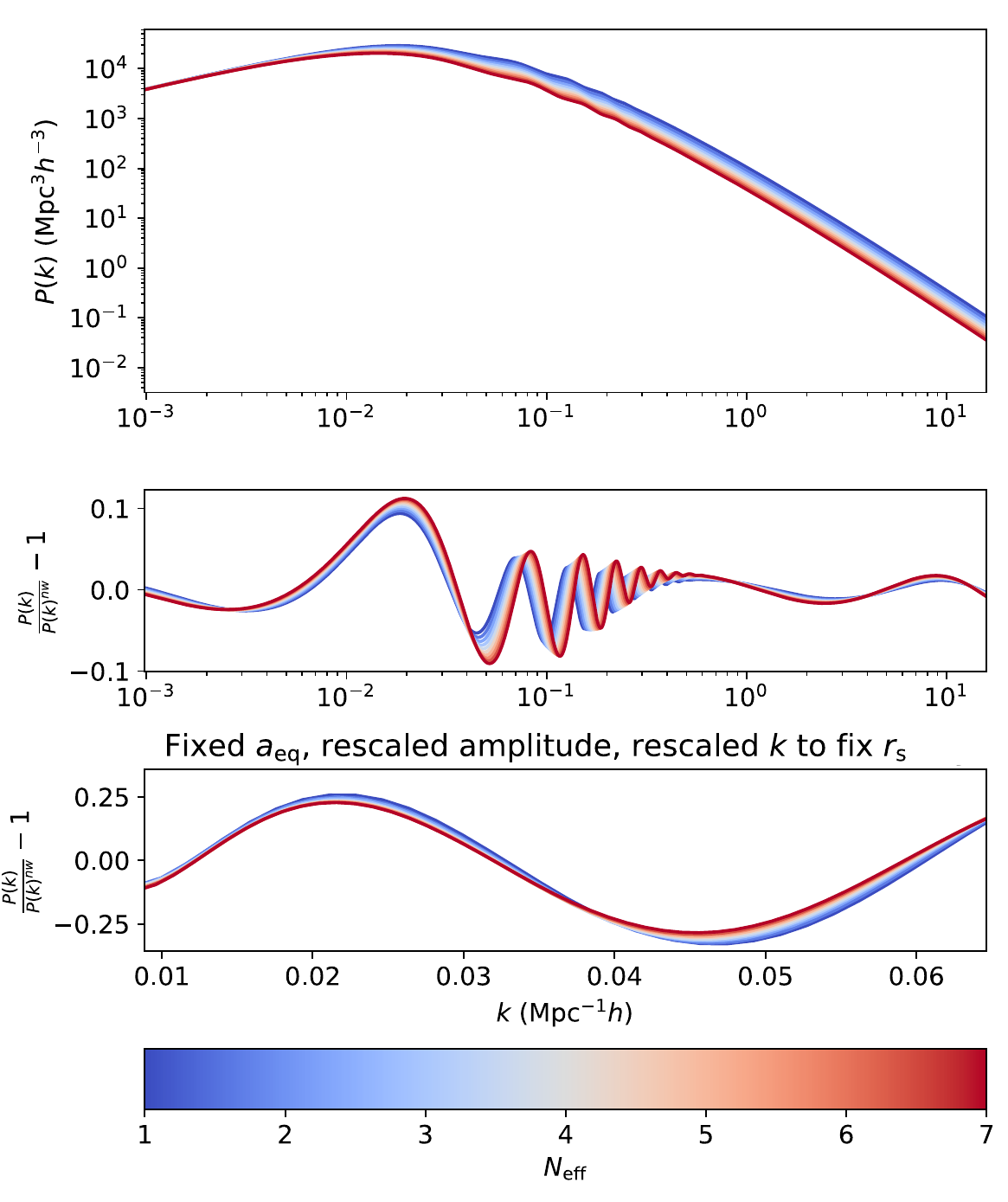}
    \caption{This figure has been inspired by Figure~3 in \protect\cite{baumann2018searching}. The top panel shows the effect of altering $N_{\mathrm{eff}}$ on the matter power spectrum, while the middle panel shows the ratio of the power spectrum plotted to a de-wiggled (smooth) power spectrum; all other cosmological parameters are held fixed. The power spectrum has been calculated using the Boltzmann solver \textsc{CLASS} \citep{blas2011cosmic}, and smoothed to isolate the wiggles using the method in \protect\cite{hinton2016measuring}. Altering $N_{\mathrm{eff}}$ has several notable effects: 1) it alters the amplitude of the BAO wiggles because the ratio of radiation to baryonic matter is changed and this affects the amplitude of rarefactions and compressions in the BAO sound waves; 2) the amplitude of the spectrum for $k$-modes that entered the horizon before the epoch of matter-radiation equality are suppressed by increasing $N_{\mathrm{eff}}$ because the additional radiation density suppresses structure growth of these modes during the era that radiation dominates the expansion rate; and 3) additional radiation alters the position of the peaks and the large turnover by increasing the expansion rate, which allows the epoch of matter-radiation equality and recombination to occur at an earlier time; the scales are smaller than in a Universe that takes longer to reach these epochs. In the lowest panel these effects have been removed to isolate the effect of $N_{\mathrm{eff}}$ on only the phase of the BAO wiggles. This is done by calculating the power spectrum wiggles as in the middle panel, but with various differences. When altering $N_{\mathrm{eff}}$ we also alter $\Omega_{\rm{m}}$ such that $a_{\mathrm{eq}} = \frac{\Omega_{\rm{r}}}{\Omega_{\rm{m}}}$ is held fixed. For each model with different $N_{\rm{eff}}$ and $\Omega_{\rm{m}}$, we rescale the $k$ modes plotted on the $x$-axis by the ratio of $r^{f}_s$ and the model $r_s$ and (where $f$ is a fiducial model with $N_{\mathrm{eff}} = 3.044$, $\Omega_{\rm{m}} = 0.315$) so that the impact of the changing $r_s$ is effectively removed. Finally, the amplitude of the wiggles is rescaled so that there is a fixed maximum height for any of the peaks shown.  }
    \label{fig::effectalteringneff}
\end{figure}

The BAO analysis can be extended to additionally account for anisotropic distortions, which are described by the Alcock-Paczynski (AP, \citealt{alcock1979evolution}) effect. In essence, the difference between the assumed fiducial cosmology model and the true cosmology distorts the distribution of galaxies in the sky differently in directions transverse or parallel to the line-of-sight \citep{bernal2020robustness}. The effect arises due to transforming the redshift-space co-ordinates of objects to real-space coordinates, and the distortions are more significant at higher redshift. In the anisotropic analysis, an additional scaling parameter $\alpha_{\mathrm{AP}}$ can be measured, representing the \textit{an}isotropic elongation or compression of the BAO. \footnote{It is also common to change variables from $\alpha$ and $\alpha_{\mathrm{AP}}$ to $\alpha_{\parallel}$ and $\alpha_{\perp}$, which represent the scaling of the BAO parallel and transverse to the line-of-sight respectively. A simple transformation exists between the two bases, which we present in Section~\ref{sec::methods}.} Depending on the constraining power of the data available, it may only be possible to measure the $\alpha$ parameter using the isotropic fitting methodology. That is, if only the monopole multipole (the first $\ell = 0$ term in the Legendre expansion of the anisotropic power spectrum corresponding to the spherically isotropic component) of the galaxy power spectrum is constrained by the data. If the quadrupole multipole ($\ell = 2$) is constrained this contains more information about the shape of the power spectrum anisotropy and enough information to measure $\alpha$ and $\alpha_{\mathrm{AP}}$ in an anisotropic fit. The fits to the values of $\alpha$ and $\alpha_{\mathrm{AP}}$ can eventually be turned into constraints on cosmological parameters or models, or combined with other datasets such as \textit{Planck} \citep{aghanim2020planck} to improve constraints on the matter density $\Omega_{m}$ and the expansion rate $H_0$. As various cosmological parameters that change the expansion rate will have degenerate effects on $\alpha$ and $\alpha_{\mathrm{AP}}$, the distortion parameters may require measurements in several redshift bins to obtain useful constraints on cosmological parameters.

The majority of analyses to this date have not taken into account an additional piece of information that the imprint of the BAOs in galaxy surveys may give us about light relics in the early Universe. \cite{bashinsky2004neutrino} first showed that free-streaming neutrinos (or any light relics that free-stream) with a speed faster than the sound speed in the early Universe induce a shift in the phase of the BAO wiggles that can be seen in Fourier space \citep[an improved semi-analytic calculation of the phase shift can also be found in][]{green2020phase}. This phase shift can be understood in that as neutrinos propagate after decoupling from the primordial plasma, they alter the gravitational potential that influences it by carrying away energy. This alteration of the potential induced a temporal shift in the BAO sound waves, that consequently led to a shift in the scale at which the sound waves froze in place and seeded the Large-scale Structure we see today. The phase shift is also discussed in detail in \cite{baumann2016phases}. Under the assumption that the primordial fluctuations are adiabatic, this phase shift is not degenerate with any other parameters and is a unique signature of neutrinos. Even if there were non-adiabatic primordial fluctuations, the scale dependence of the phase shift may still possibly allow it to be distinguished from non-adiabatic fluctuations \citep{baumann2018searching}. 

\begin{figure}
    \centering \includegraphics[width=0.47\textwidth]{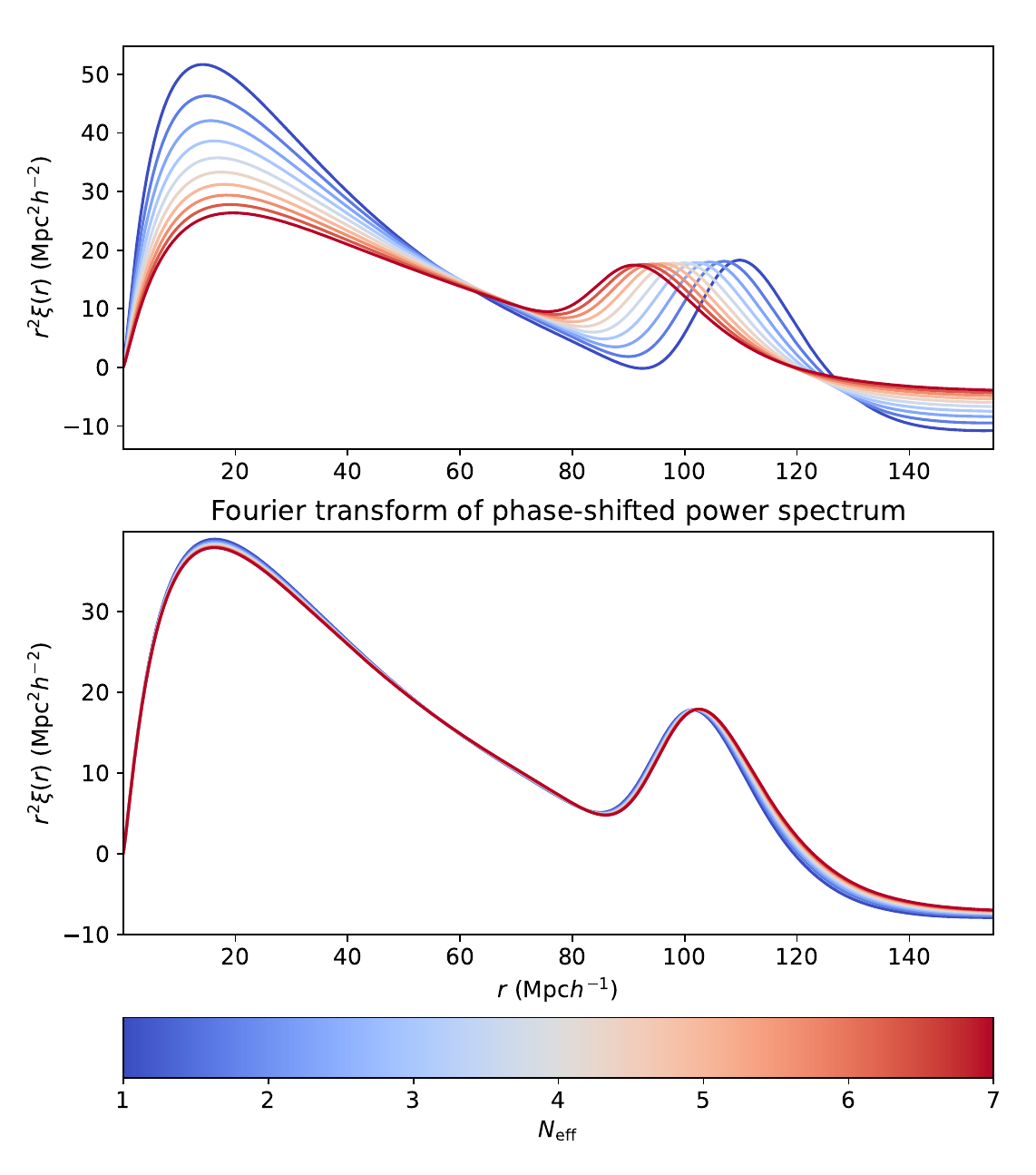}
    \caption{\small Top panel: the effect of altering $N_{\mathrm{eff}}$ on the correlation function, calculated from a Fourier transform of the power spectrum shown in the top panel of Figure~\ref{fig::effectalteringneff}. Lower panel: the correlation function calculated from the power spectrum, using a Fourier transform, after including a phase shift following the parameterization of the phase shift shown in \protect\cite{baumann2019first}. The power spectrum is shifted along the $x$-axis by shifting the $k$ values by an amount expected by for a given $N_{\mathrm{eff}}$ and interpolating the power spectrum. Unlike the top panel, the lower panel shows only the effects due to a phase shift on the correlation function rather than the various impacts that can be seen in the top panel due to $N_{\mathrm{eff}}$, including the phase shift. }
    \label{fig::effectalteringneffcorr}
\end{figure}

The phase shift was first parameterized and measured in CMB data from the \textit{Planck} satellite by \cite{follin2015first}, whose analysis showed a probability $p = 8 \times 10^{-6}$ for the absence of a phase shift due to free-streaming relics, indicating a strong presence of free-streaming relics. More recently, it was parameterized and measured in BAO data from BOSS DR12 by \cite{baumann2019first} who found with 95\% confidence a non-zero phase shift by free-streaming relics, and additionally in \cite{montefalcone2025free} with CMB data. The parameterization by \cite{baumann2019first}, like the standard BAO analysis, involves a semi model-independent parameter $\beta$ that characterizes the amplitude of the phaseshift and can be transformed into a constraint on $N_{\mathrm{eff}}$, the effective number of neutrino species. The effect of altering $N_{\mathrm{eff}}$ in the matter power spectrum, which shifts the phase in the BAOs, can be seen in Figure~\ref{fig::effectalteringneff}. The effect in configuration space can be seen in Figure~\ref{fig::effectalteringneffcorr}, where the correlation function has been calculated from the power spectrum using the \emph{mcfit} package.\footnote{\url{https://github.com/eelregit/mcfit}} It is apparent from the figures, that the impact of the phase shift from varying $N_{\mathrm{eff}}$ on the BAO wiggles is subtle compared to various other impacts $N_{\mathrm{eff}}$ has on the power spectrum or correlation function. Nonetheless, it can be detected. 

\cite{baumann2018searching, baumann2019first} also discuss and provide forecasts for the potential for upcoming galaxy surveys that will probe the Large-scale Structure to constrain the phase shift and $N_{\mathrm{eff}}$. One such survey is the DESI survey \citep{aghamousa2016desi, aghamousa2016desi2, abareshi2022overview, adame2024validation}, a 5-year galaxy survey that is currently in progress and began in 2020.\footnote{\url{https://www.desi.lbl.gov/}} It is using an instrument that consists of 5000 robot fibres, mounted on the focal place of the Mayall Telescope at Kitt Peak, Arizona, to measure spectra for 35 million galaxies. The DESI instrument has already detected the BAO signal in their DR1 Data \citep{adame2024desi3} with a $9.1 \sigma$ level detection at 0.86\% precision using a sample of luminous red galaxies (LRGs) and emission line galaxies (ELGs). In \cite{baumann2018searching, baumann2019first}, it was anticipated using Fisher Matrix Forecasts that DESI will be able to measure the phase shift with an uncertainty of $\sigma(\beta) \sim 0.3$ from their 5-year survey; DESI BAOs + \textit{Planck} may achieve $\sigma(N_{\mathrm{eff}}) = 0.15$, and with the full broadband power spectrum from DESI may achieve $\sigma(N_{\mathrm{eff}}) = 0.087$.

In this paper, we start down this path by constraining the phase shift with DESI DR1 BAO data. Unlike the constraints shown in \cite{baumann2019first}, we also extend the methodology to constrain $\beta$ with both the isotropic and anisotropic BAO parameters, $\alpha$ and $\alpha_{\mathrm{AP}}$. In Section~\ref{sec::methods}, we will give an overview of the anisotropic fitting methodology and the parameterization of the phase shift given by \cite{baumann2019first} and our implementation of the method in two BAO fitting codes; \textit{Barry} \citep{hinton2020barry} and \textit{desilike}.\footnote{\url{https://desilike.readthedocs.io/en/latest/\#}} In Section~\ref{sec::validationmocks}, we will show the robustness of the approach and codes applied to mocks produced for the DESI collaboration for datasets including the Bright Galaxy Survey (BGS) \citep{hahn2023desi}, LRGs \citep{zhou2023target}, ELGs \citep{raichoor2023target} and quasars (QSOs) \citep{chaussidon2023target}, and constraints for the combinations of these datasets. Our results from applying the methodology to the DESI DR1 BAO data will be shown in Section~\ref{sec::resultswithyr1data} and we conclude in Section~\ref{sec::conclusions}.

\section{Methodology}\label{sec::methods}

In this work we follow the methodology of \cite{baumann2018searching, baumann2019first} to constrain the phase shift. As mentioned previously, instead of only constraining the isotropic BAO scaling parameter $\alpha$ and the phase shift, we extend the approach to measure the anisotropic BAO by fitting $\alpha_{\parallel}$, $\alpha_{\perp}$ and the phase shift amplitude $\beta$. Furthermore, we implement this methodology in two codes to fit these parameters; \textit{Barry} and \textit{desilike} which can both fit the BAO wiggles using the galaxy power spectrum or the BAO peak in the correlation function. Having these two approaches will allow us to see the robustness of our constraint under different methodologies.

\subsection{Isotropic BAO analysis}

In the standard isotropic BAO analysis, the galaxy power spectrum can be split into a smoothed component (or no-wiggle component/broadband component) and a wiggle component,

\begin{align}\label{eq::powerspectrumwiggledecomp}
    P(k) &= P_{\mathrm{nw}}(k) + P_{\mathrm{w}}(k) \\ \nonumber
         &= P_{\mathrm{nw}}(k)\left(\frac{P_{\mathrm{w}}(k)}{P_{\mathrm{nw}}(k)} + 1\right) \\ \nonumber 
         &= P_{\mathrm{nw}}(k) \left( \mathcal{O}(k) + 1 \right).
\end{align}
The goal of the analysis is to isolate the BAO wiggles from the observed power spectrum, and fit the distortion parameter $\alpha$ which scales the true positions of the BAO peaks and troughs in Fourier space. The distance scaling of the peaks that occurs is by an amount relative to the expected positions of the BAO peaks in a \emph{fiducial} model that is assumed when the observed redshifts of galaxies are transformed to distances, and also relative to a \emph{template} cosmological model that is assumed when one is fitting the BAOs; often these fiducial and template cosmologies are the same. However in general, the parameter $\alpha$ can be defined as 
\begin{equation}
    \alpha = \frac{ D_V(z) }{ D^{\mathrm{f}}_V(z) } \frac{r^{\mathrm{t}}_s}{r_s}.
\end{equation}
$D_V(z)$ corresponds to the volume averaged distance,
\begin{equation}
    D_V(z) = \left( (1+z)^2 D_A^2(z) \frac{cz}{H(z)} \right)^{1/3},
\end{equation}
where $H(z)$ is the Hubble parameter at the redshift $z$ corresponding to the redshift of the BAO measurement, and $D_A(z)$ is the angular diameter distance. $r_s$ corresponds to the sound horizon at the baryon drag epoch (at $z_{d}$). The quantities with the `f' superscript refer to the expected value in the fiducial model, while those with the `t' superscript refer those in the template model. When all of these models are the same as the true cosmology, $\alpha = 1$; but deviations from unity allow us to quantify the deviation of the models from the true cosmology. $\alpha$ enters the power spectrum as a rescaling of the measured $k$-modes, and one writes the measured modes $k' = k/\alpha$, where $k$ denotes the true modes. The power spectrum is also scaled by a ratio of the volume averaged distances cubed due to the fact that the rescaling of distances changes the defined survey volume, which must be accounted for as the power spectrum has units of volume. Typically the redshift-space power spectrum $P(k,\mu)$ is decomposed into moments using a Legendre polynomial expansion $(\mathcal{P}_{\ell}(\mu))$ as
\begin{equation}\label{eq::powerspectrummoments}
    P_{\ell}(k) = \frac{2\ell + 1}{2} \int_{-1}^{1} d\mu P(k,\mu) \mathcal{P}_{\ell}(\mu),
\end{equation}
allowing us to write the following expression for the redshift-space power spectrum as a summation over the power spectrum moments, 
\begin{equation}
    P(k,\mu) = \sum_{\ell} P_{\ell}(k)\mathcal{P}_{\ell}(\mu).
\end{equation}
The isotropic analysis allows us to constrain the monopole moment corresponding to $\ell = 0$. Here, $\mu = \cos{(\mathbf{k}\cdot \mathbf{\hat{r}})}$ -- the cosine of the angle between the wavevector of interest and the line-of-sight. The parameter $\alpha$ enters the wiggle power spectrum template as 
\begin{equation}
    P_{\mathrm{w}}'(k) = P^{\mathrm{t}}_{\mathrm{w}}(k/\alpha).
\end{equation}
 
\subsection{Anisotropic BAO analysis}

In the anisotropic case, it is possible to constrain both the monopole and the quadrupole corresponding to $\ell = 2$. The analysis constrains two distortion parameters, $\alpha_{\parallel}$ that scales the BAO peaks along the line-of-sight while $\alpha_{\perp}$ constrains the BAO peaks transverse to the line-of-sight; when the fiducial cosmology and template cosmology agree with the real cosmology, these parameters are unity. Alternatively, the anistropic analysis can be parameterized in terms of $\alpha$ defined previously and $\alpha_{\mathrm{AP}}$ for the distortions, where $\alpha_{\mathrm{AP}}$ is also unity when all three cosmologies are the same. The parameters $\alpha_{\parallel}$ and $\alpha_{\perp}$ are defined as\footnote{Strictly, the geometric AP Effect actually is only due to distortions from the ratios $q_{\parallel} = \frac{H^{\rm{f}}(z)}{H(z)}$ and $q_{\perp} = \frac{D_A(z)}{D^{\rm{f}}_A(z)}$, because the factor of $\frac{r^{\rm{t}}_s}{r_s}$ only enters due to the choice of BAO template. However it is common to refer to $\alpha_{\parallel}$ and $\alpha_{\rm{AP}}$ in the literature when constraining the BAO distortion parameters. } 
\begin{align}\label{eq::alpha_para}
    \alpha_{\parallel} & = \frac{H^{\rm{f}}(z) r^{\rm{t}}_s}{H(z) r_s}, \\ 
    \label{eq::alpha_perp}
    \alpha_{\perp} & = \frac{D_A(z) r^{\rm{t}}_s}{D^{\rm{f}}_A(z) r_s}. 
\end{align}
Furthermore, equation~\ref{eq::alpha_para} and equation~\ref{eq::alpha_perp} can be related to $\alpha$ and $\alpha_{\mathrm{AP}}$ as 
\begin{align}
    \alpha & = \alpha_{\rm iso} =  \alpha^{1/3}_{\parallel} \alpha^{2/3}_{\perp}, \\ \nonumber
    % \epsilon & = \left( \frac{\alpha_{\parallel}}{\alpha_{\perp}} \right)^{1/3} - 1.
    \alpha_{\rm AP} & = \alpha_{\parallel}/\alpha_{\perp}.
\end{align}
If one assumes a flat $\Lambda$CDM cosmology, it is possible to directly relate $\alpha_{\mathrm{AP}}$ to the matter density $\Omega_{\rm{m}}$, 
\begin{align}
    \alpha_{\mathrm{AP}} & = \alpha_{\mathrm{AP}}(z,\Omega_{\mathrm{m}}), \nonumber \\  & = \left( H(z) D_A(z) \right)^{\mathrm{f}}  \left( H(z, \Omega_{\mathrm{m}}) D_A(z, \Omega_{\mathrm{m}} ) \right)^{-1}, \nonumber \\ 
    & = \left( E(z) D_A(z)h \right)^{\mathrm{f}} \left( E(z, \Omega_{\mathrm{m}}) D_A(z, \Omega_{\mathrm{m}} )h \right)^{-1}.
    \label{eq:alphaaptocosmo}
\end{align} 
where we have used $H(z) = 100\,\mathrm{km\,s^{-1}\,Mpc}^{-1}\, h\,E(z)$. The term containing the angular diameter distance, $D_A(z, \Omega_{\mathrm{m}} )h$, depends on only $\Omega_{\mathrm{m}}$ and $z$. $\alpha$ on the other hand relates to both $\Omega_{m}$ and the comoving sound horizon $r_sh$,
\begin{align}
    \alpha & = \alpha (\Omega_{\mathrm{m}}, r_sh), \nonumber \\ 
    & = \left[ \left( \frac{H(z)}{D^2_A(z)} \right)^{1/3} r_s^{\rm{t}} \right] ^{\mathrm{f}} \left[ \left( \frac{H(z)}{D^2_A(z)} \right)^{1/3} r_s \right]^{-1}, \nonumber \\ 
    & = \left[ \left( \frac{E(z)}{D^2_A(z)h^2} \right)^{1/3} r_s^{\rm{t}}h \right] ^{\mathrm{f}} \left[ \left( \frac{E(z)}{D^2_A(z)h^2} \right)^{1/3} r_sh \right]^{-1}.
    \label{eq:alphatocosmo}
\end{align}

It is therefore convenient to note that, for a flat $\Lambda$CDM cosmology, one can uniquely map from a set of BAO parameters to a cosmological model by first computing $\Omega_{\rm{m}}$ given $\alpha_{\mathrm{AP}}$, then using this fixed value alongside $\alpha$ to isolate $r_sh$. The distortion parameters enter the wiggle power spectrum template as 
\begin{align}
    P_{\mathrm{w}}(k,\mu) & = P^{\rm{t}}(k(k',\mu'), \mu(k',\mu')), \\ \nonumber 
    & = P^{\rm{t}}(k(\alpha_{\parallel},\alpha_{\perp}), \mu(\alpha_{\parallel},\alpha_{\perp})),
\end{align}
where $k'$ and $\mu'$ are the observed (distorted) $k$ and $\mu$ due to anisotropic distortions. The observed $k'$, $\mu'$ can be related to the real $k$, $\mu$ as 
\begin{align}
    k' & =  \frac{k}{q_{\perp}}\left[ 1 + \mu^2( \frac{1}{F^2}- 1) \right]^{1/2}, \\ \nonumber 
    \mu' & = \frac{\mu}{F}\left[ 1 + \mu^2( \frac{1}{F^2}- 1) \right]^{-1/2},
\end{align}
where $F = q_{\parallel}/q_{\perp}$, $q_{\parallel} = H^{\rm{f}}(z)/H(z)$ and $q_{\perp} = D_A(z)/D_A{\rm{f}}(z)$.

\subsection{Anisotropic BAO analysis with \texorpdfstring{$\beta_{\phi}$}{TEXT}}

Similar to the BAO distortion parameters, the phase shift is constrained by measuring the difference in the amplitude of a phase shift from a template and the true cosmology. However, one should keep in mind that while the phase shift is a parameter similar to $\alpha$ and $\alpha_{\rm AP}$ in that it is model independent and can be transformed into a constraint on cosmology, it is not a geometric distortion but is a physical change to the true $k$ modes in the power spectrum. The BAO wiggle feature can be expressed as a sinusoidal oscillation with a $k$-dependent amplitude,
\begin{equation}
    O_{\mathrm{lin}}(k) \propto A(k) \sin{(k r_s + \phi(k))},
\end{equation}
where $\phi$ is the phase shift; it is parameterized in \cite{baumann2018searching, baumann2019first} as 
\begin{equation}
    \phi(N_{\mathrm{eff}},k) = \beta_{\phi}(N_{\mathrm{eff}}) f(k).
\end{equation}
Here $\beta_{\phi}$ is the relative amplitude of the phase shift compared to a template model, and $f(k)$ is a fitting function for the scale dependence of the $\phi$. When the template cosmology model chosen and the true cosmology match, $\beta_{\phi}$ is unity. $\beta_{\phi}$ is related to $N_{\mathrm{eff}}$ as 
\begin{equation}
    \beta_{\phi} = \frac{\epsilon_{\nu}}{\epsilon^{\rm{t}}_{\nu}} = \frac{N_{\mathrm{eff}}/(N_{\mathrm{eff}} + a_{\nu})}{ N^t_{\mathrm{eff}}/(N^{\rm{t}}_{\mathrm{eff}} + a_{\nu}) }; \qquad
    a_{\nu} = \frac{8}{7}\left(\frac{11}{4}\right)^{4/3}.
\end{equation}
Given that the phase shift is a physical effect caused by free-streaming relativistic species, rather than an effect that is induced by the choice of the fiducial model used to construct the power spectrum or correlation function, we expect $\beta_{\phi}$ to be independent of the choice of fiducial cosmology and only dependent on the template cosmology. If $N_{\mathrm{eff}} = 0$, we also have $\beta_{\phi} = 0$. In the limit $N_{\mathrm{eff}} \rightarrow \infty$, $\beta_{\phi} \approx 2.4$ when $N^t_{\mathrm{eff}} = 3.044$. $f(k)$ is redshift independent and has been approximated by \cite{baumann2018searching}; it is to first order independent of $N_{\mathrm{eff}}$ for a range of values of $N_{\mathrm{eff}}$. This can be seen in Figure~1 of \cite{baumann2019first}. The fitting function template $f(k)$ is parameterized as 
\begin{equation}\label{eq::phaseshift_kdependence}
    f(k) = \frac{\phi_{\infty}}{1 + (k_*/k)^{\xi}}, 
\end{equation}
with $\phi_{\infty} = 0.227$, $k_* = 0.0324 h\mathrm{Mpc}^{-1}$ and $\xi = 0.872$. As $k \rightarrow \infty$, $f(k) \rightarrow \phi_{\infty}$. Finally, the observed wiggle power spectrum can be written in terms of the anisotropic distortion parameters and the phase shift amplitude as 
\begin{align}\label{eq::phaseshiftimplementationWIGGLES}
    P_{\mathrm{w}}(k,\mu) & = P^{\rm{t}}(k(k',\mu') + (\beta_{\phi} - 1)f(k')/r_s, \mu(k',\mu')), \\ \nonumber 
    & = P^{\rm{t}}(k(\alpha_{\parallel},\alpha_{\perp}) + (\beta_{\phi} - 1)f(k')/r_s, \mu(\alpha_{\parallel},\alpha_{\perp})).
\end{align}
It is also possible to directly map the anisotropic BAO distortion parameters to $\Omega_{\rm{m}}$ and $r_sh$ in a flat $\Lambda$CDM cosmology. We note here that unlike equation~3.4 in \cite{baumann2019first} we include the impact of the distortions in the scale-dependent function $f(k)$ for the phase shift in equation~\ref{eq::phaseshiftimplementationWIGGLES}. However, we found there was no impact on the fits to any parameters by choosing to include the distortion $k'$ rather than simply $k$ in $f(k)$ (which is how the model is implemented in \emph{Barry}); however the impact may be more significant in cases one explores exotic models which alter the function $f(k)$.

\subsection{Implementation}

Our fitting methodology for the BAO scaling parameters and phase shift amplitude has been implemented in both \emph{desilike}\footnote{\url{https://github.com/cosmodesi/desilike}} and \emph{Barry}\footnote{\url{https://barry.readthedocs.io/en/latest/}} \citep{hinton2020barry}. These implementations are both \textit{python} packages written for fitting the BAO feature and obtaining constraints on the BAO parameters. While both codes are capable of standard isotropic and anisotropic fits \footnote{the \emph{Barry} code has been updated since the work presented in \cite{hinton2020barry}, see more about the improvements in \cite{chen2024baryon}}, in this work we have extended \emph{Barry} in order to additionally constrain the phase shift.\footnote{\url{https://github.com/abbew25/Barry}} \emph{desilike} also contains the implementation via recent updates. While these codes have various available clustering and BAO models to choose from, here we focus on using and implementing the phase shift parameterisation with a model similar to the style of \cite{beutler2017clustering}. This model has been widely used in previous BAO analyses and recently by the DESI collaboration for the analysis of the year 1 data \citep[the model used is presented and discussed in][]{chen2024baryon}. We use the same modelling to be consistent with the DESI Year 1 results \citep{adame2024desi}. 

\subsubsection{Power spectrum}

In \emph{desilike} and \emph{Barry} the redshift-space power spectrum is generically expressed in the following way for the BAO analysis, using the decomposition of the power spectrum into a broadband (no-wiggles) component and a sinusoidal wiggles component as in equation~\ref{eq::powerspectrumwiggledecomp} \citep{chen2024baryon},
\begin{equation}
    P(k,\mu) = \mathcal{B}(k,\mu) P_{\mathrm{nw}}(k) + \mathcal{C}(k,\mu) P_{\mathrm{w}}(k)  + \mathcal{D}(k,\mu).
\end{equation}
$\mathcal{B}$ accounts for peculiar velocity effects such as the Kaiser Effect \citep{kaiser1987clustering} and the Finger-of-God Effect \citep[FOG,][]{jackson1972critique}, $\mathcal{C}$ is the propagator which accounts for non-linear evolution of the BAO feature, and $\mathcal{D}$ accounts for additional features in the measured broadband component of the power spectrum not captured by the no-wiggle component, for instance due to systematics or non-linear structure growth. This term is discussed in more detail in sections~\ref{subsec::polynomialbroadbandmethod} and \ref{subsec::splinebroadbandmethod}.
In our model, we express $\mathcal{B}$ as 
\begin{equation}
    \mathcal{B}(k,\mu) = (b + f\mu^2)^2\left[ 1 + \frac{1}{2} k^2 \Sigma_s^2 \right]^{-2}
\end{equation}
where $b$ represents the linear galaxy bias parameter, $f$ is the growth rate of structure (distinctly different from the fitting function for the neutrino phase shift, $f(k)$), and where both parameters combine to form the Kaiser factor $(b + f\mu^2)^2$. $\Sigma_s$ accounts for FoG damping, and all three of these are free parameters in the BAO analysis. The propagator is expressed as 
\begin{equation}
    \mathcal{C} = (b + f\mu^2)^2 \exp{\left[ -\frac{1}{2}k^2(\mu^2 \Sigma_{\mathrm{nl,\parallel}}^2 + (1 - \mu^2) \Sigma_{\mathrm{nl,\perp}}^2) \right]}
\end{equation}
where $\Sigma_{\mathrm{nl,\parallel}}$, $\Sigma_{\mathrm{nl,\perp}}$ are additional free parameters that account for non-linear damping of the BAO due to bulk motions of galaxies (typically on scales $\sim 10 h^{-1}$ Mpc.) separately for damping along and perpendicular to the line-of-sight respectively. For an isotropic analysis this simply reduces to 
\begin{equation}
    \mathcal{C} = (b + f\mu^2)^2 \exp{\left[ -\frac{1}{2}k^2 \Sigma_{\mathrm{nl}}^2 \right]},
\end{equation}
involving only one isotropic damping parameter $\Sigma_{\mathrm{nl}}$. 

\subsubsection{Reconstruction}

In a BAO analysis it is typical to apply an type of algorithm called \emph{reconstruction} to the data, which enhances the BAO signal allowing for improved signal-to-noise. Data products, such as the power spectrum, where a reconstruction algorithm has been applied are referred to as \emph{post-recon} while those without the application of this algorithm are called \emph{pre-recon}. This algorithm is described in greater detail theoretically in \cite{chen2024baryon} and algorithms were tested practically with optimal settings for DESI data found in \cite{paillas2024optimal, chen2024extensive}, however the basic steps are described here. First, it involves smoothing the galaxy density field with a Gaussian filter to filter out more non-linear modes. Then, the Zeldovich approximation is used in order to calculate the displacement of the positions of galaxies in space due to structure growth, and this is used to shift the positions of a random catalogue of galaxies and the galaxies in the data by subtracting the calculated linear theory displacement. The reconstructed density field is then given by the difference between the shifted galaxy density field of the data and the shifted random catalogue. The effect of this is to sharpen the BAO feature in the data, because distortions due to non-linear structure growth and the bulk motions of galaxies are largely removed. Consequently, the parameters $\Sigma_{\mathrm{nl,\parallel}}$, $\Sigma_{\mathrm{nl,\perp}}$ and $\Sigma_s$ vary between mocks and data that are pre-recon and post-recon, with reconstruction reducing the values of the damping parameters.

In this work we use post-recon mocks that have been reconstructed using the RecSym convention in DESI, in which case the displacement applied to the random catalogue and data are equivalent and the anisotropy in redshift space is preserved \citep[this in contrast to the RecIso formalism -- a more detailed discussion of both can also be seen in][]{chen2024baryon,paillas2024optimal}. 

\subsubsection{Polynomial broadband methodology}\label{subsec::polynomialbroadbandmethod}

Historically, the term $\mathcal{D}(k,\mu)$ has been modelled using a polynomial expansion. In the methodology of \cite{beutler2017clustering}, this can be expressed for the monopole and quadrupole for the pre-recon data as 
\begin{equation}
    \mathcal{D}_{\ell}(k) = \frac{a_{{\ell},1}}{k^3} +  \frac{a_{{\ell},2}}{k^2} + \frac{a_{{\ell},3}}{k} + a_{{\ell},4} + a_{{\ell},5}k.
\end{equation}
For the post-recon analysis, it is written as 
\begin{equation}
    \mathcal{D}_{\ell}(k) = \frac{a_{{\ell},1}}{k^3} +  \frac{a_{{\ell},2}}{k^2} + \frac{a_{{\ell},3}}{k} + a_{{\ell},4} + a_{{\ell},5}k^2.
\end{equation}
The coefficients $a_{{\ell},i}$ and number of polynomial terms are chosen based on the best fit to the data. To be consistent with the DESI methodology and formalism shown in \cite{chen2024baryon} and \cite{paillas2024optimal}, we fit for all the coefficients in our analysis. 

\subsubsection{Spline broadband methodology}\label{subsec::splinebroadbandmethod}

The polynomial broadband methodology suffers from some limitations \citep[this is described in detail in][]{chen2024baryon} due to the fact it is not always clear what the optimal number or form of polynomial coefficients to use should be, and when transformed to configuration space they may result in a peak (mimicking the BAO peak). Therefore \cite{chen2024baryon} present a new approach to modelling additional contributions to the broadband power spectrum, using a summation of cubic splines with bases separated by some choice $\Delta$ (set to 0.06 $h^{-1}$ Mpc in our work),
\begin{equation}
    \mathcal{D}_{\ell}(k) = \sum_{n=-1}^{\infty} a_{{\ell},n} W_3 \left(\frac{k}{\Delta} - n \right).
\end{equation}

$W_3(k)$ is a fourth order piece-wise cubic spline that is given in \cite{chen2024baryon}. This approach importantly removes degeneracy between the broadband component of the power spectrum and the BAO peak and reduces the choice of both the form and number of polynomials to a single choice of $\Delta$ (which has a well-defined lower bound --- it should be larger than half the BAO wavelength). This choice of $\Delta$ alongside the range of scales being fit automatically defines the number of free broadband parameters. Distinct terms can also be set for each multipole moment of the power spectrum. In the analysis our choice of fitting scales ($k = 0.02-0.30$ $\rm{Mpc}^{-1} h$) and $\Delta$ give 7 $a_{{\ell},n}$ coefficients to be fit for each multipole. In our analysis, we test the performance of both the spline methodology and the polynomial methodology in our fits to the phase shift. 

\subsubsection{Correlation function}\label{sec::corrfunc_splinebroadbandconfig}

To analyse the BAO in configuration space, the redshift-space correlation function is simply calculated from a Hankel transform of the redshift-space power spectrum to configuration space. The only difference in our analysis of the correlation function compared to the power spectrum is that we set the number of coefficients to only three for analyses using the polynomial broadband methodology. For the spline broadband methodology, the transformation of the cubic spline basis presented above results in two large scale $n=[0,1]$ terms for the quadrupole which we include, while the remainder are confined to small scales outside our fitting range ($50-150$ $\rm{Mpc}h^{-1}$). In addition, the scale-cuts used when fitting the power spectrum to avoid systematics transform into simple polynomials. As such, the final form of $\mathcal{D}_{{\ell}}(k)$ for spline-based correlation function fits is completely analogous to that for the power spectrum and takes the form
\begin{align}
    \mathcal{D}_{0}(k) & = a_{0,0} + a_{0,1} \left( \frac{rk_{\mathrm{min}}}{2\pi} \right)^2, \\ \nonumber 
    \mathcal{D}_{2}(k) & = a_{0,0} + a_{0,1} \left( \frac{rk_{\mathrm{min}}}{2\pi} \right)^2 + a_{2,0} B_{2,0}(r\Delta) + a_{2,1} B_{2,1}(r\Delta),
\end{align}
where 
\begin{equation}
    B_{{\ell},n}(r\Delta) = i^2 \int \frac{dk k^2}{2 \pi^2 } W_3 \left(\frac{k}{\Delta} - n \right) j_{\ell}(kr),
\end{equation}
is the result of transforming the cubic spline functions and $k_{\mathrm{min}}$ refers to the minimum choice of $k$ used in our Fourier-space fits. $j_{\ell}(kr)$ are the spherical Bessel functions of the first kind. The reconstruction procedure using RecSym convention is also applied in the case of the correlation function analysis. 

\section{Mock validation}\label{sec::validationmocks}

In order to validate the methodology we first apply the fitting procedure in \emph{Barry} and \emph{desilike} to various mocks for the DESI data. The DESI survey strategically splits the survey into dark-time and bright-time components, to optimize detection of galaxies when the moon illuminates the sky (bright-time, so that only bright galaxies at low-redshift can be detected) and dark-time for fainter objects. As such, the DR1 data consists of BGS (galaxies detected under the bright-conditions), LRGs, ELGs and QSOs \citep[see more detail in][]{adame2024desi}. Mocks are provided for the individual datasets of galaxies detected in the DR1 data set, split into various redshift bins as required, and we obtain fits for each of these individual bins. We also provide combined fits to the mocks by combining the likelihoods for each of the fits to the individual bins. While we produce fits with the $\beta_{\phi}$ parameter using the anisotropic BAO methodology for the majority of the mocks and corresponding data, to be consistent and comparable to the results provided by the DESI Collaboration for the DR1 data \citep{adame2024desi} we only fit $\beta_{\phi}$ using an isotropic BAO procedure for QSOs, BGS and for ELGs at $z=0.8-1.1$. Furthermore, we fit a combined measurement of the correlation function and power spectrum for ELGs and LRGs at $z=0.8-1.1$ to account for their cross-correlation. Otherwise, following the DESI Collaboration, we neglect cross-correlations between QSOs with ELGs and LRGs from $z=0.8-1.1$ due to the fact that the number of overlapping tracers in the survey volume is likely to be negligible, and neglect cross-correlations between other tracers which are detected in non-overlapping redshift bins. The details of the survey tracers are summarised in the Table~\ref{tab:summarydata}. 

\begin{table}
    \centering
    \caption{\small Summary of DESI DR1 data tracers properties. The LRGs are split across three redshift bins (LRG1, LRG2, LRG3) and ELGs across two (ELG1, ELG2). }
    \begin{tabular}{>{\raggedright\arraybackslash}p{1.5cm} S[table-format=1.5] >{\centering\arraybackslash}p{1.8cm}}
       \toprule
       \midrule
       \textbf{Tracer} & {$z$} & {$N$ objects} \\ \midrule
       BGS & {0.1 - 0.4} & {300,017} \\ \midrule
       LRG1 & {0.4 - 0.6} & \\[1.5mm]
       LRG2 & {0.6 - 0.8} & {2,138,600} \\[1.5mm]
       LRG3 & {0.8 - 1.1} & \\ \midrule
       ELG1 & {0.8 - 1.1} & \multirow{2}{*}{\makecell{2,432,022}} \\[1.5mm]
       
       ELG2 & {1.1 - 1.6} & \\ \midrule
       QSO & {0.8 - 2.1} & {856,652} \\
       \bottomrule
    \end{tabular}
    \label{tab:summarydata}
\end{table}

\subsection{Fitting DESI first-generation mocks}

\subsubsection{Description of first-generation mocks}

In this paper, DESI first-generation mocks \citep{adame2024desi3, adame2024desi} refer to a set of single-redshift snapshots of full N-body mocks based on the Abacus Summit simulations \citep{abacussummitsims21}.\footnote{\url{https://abacussummit.readthedocs.io/en/latest/abacussummit.html}} The first-generation mocks are cubic boxes with a \textit{Planck} 2018 \citep{aghanim2020planck} cosmology (called c000 cosmology in this paper, with $N_{\mathrm{eff}} = 3.044$), used here in order to test the BAO fitting methods on mocks with precision far beyond that of DESI DR1. The dark matter haloes are populated in the mocks using a Halo Occupation Distribution algorithm \citep[HOD,][]{rocher2023desi, yuan2024desi}. We also make use, where available, of first-generation `control variate' (CV) mocks, which have been processed to reduce the noise in the galaxy clustering signal by subtracting off the intrinsic sample variance in the simulated observables \citep{hadzhiyska2023mitigating}. These allow us to test the methodologies with greater precision, and quantify the effects of systematics when the statistical noise in the mocks is smaller than we expect to see even for the final DESI data. More details about the mocks can also be found in \cite{garcia2024hod, mena2024hod}. The `second-generation' mocks discussed later capture the selection function of the DR1 data and involve an improved HOD algorithm compared to the first-generation mocks. The first-generation mocks do not include systematics in the DESI pipeline like in second-generation mocks. The covariance matrices for the mocks have been produced using 1000 realisations of EZmocks \citep{chuang2015ezmocks}, which are produced with the same selection function as is used for the Abacus Summit mocks. We apply the Hartlap correction to the covariance matrices for these mocks \citep{hartlap2007your, percival2014clustering, percival2022}. 

For all the fits to the first-generation mocks, to be consistent with the modelling in \cite{chen2024baryon}, a Gaussian prior $\mathcal{N}(\mu,\sigma)$ with mean $\mu$ and width $\sigma$ has been used for $\Sigma_s$ given by $\mathcal{N}(2,2)$ ($\mathrm{Mpc} h^{-1}$), for $\Sigma_{\mathrm{nl, \parallel}}$ given by $\mathcal{N}(5, 2)$ ($\mathrm{Mpc} h^{-1}$) and $\Sigma_{\mathrm{nl,\perp}}$ given by $\mathcal{N}(2,1)$ ($\mathrm{Mpc} h^{-1}$). We have allowed $\beta_{\phi}$ to vary over $\mathcal{U}(-1, 3)$ (uniformly over the range $[-1,3]$) as a minimum range. For polynomial fits, three and five terms with power-law indices of $[-2, -1, 0]$ and $[-1, 0, 1, 2, 3]$ are used to fit the broadband signal for the correlation function and power spectrum respectively. For the spline methodology, seven coefficients are used for the cubic splines and two coefficients are used for the spline basis which is Hankel transformed for the correlation function fits (in addition to two more coefficients for polynomial terms -- the details can be seen in Section~\ref{subsec::splinebroadbandmethod}). We let $\alpha$ vary over $\mathcal{U}(0.8, 1.2)$ (uniformly over the range [0.8, 1.2]) and the parameter $\epsilon = \alpha_{\mathrm{AP}}^{1/3}-1 $ varies over $\mathcal{U}(-0.2, 0.2)$ in \textit{Barry} ($\alpha_{\mathrm{AP}}$ varies over $\mathcal{U}(-0.2, 0.2)$ in \textit{desilike}).

\subsubsection{Comparison of \emph{desilike} with Baumann et al pipeline}

In order to validate the methodology implemented in \emph{desilike} (using the \emph{emcee} sampler, \citealt{foreman2013emcee}) we begin by fitting the post-recon correlation function of 25 CV ELG mocks at $z = 1.1$ using both this code and the isotropic pipeline of \cite{baumann2019first}. The results, which can be seen in Appendix~\ref{app::fitsbaumannvalidation}, indicate the implemented scheme reasonably fits the BAO and phase shift parameters, and we see the expected correlation between $\alpha$ and $\beta_{\phi}$. We expect that by including $\alpha_{\rm{AP}}$ as an additional parameter, this should have little impact on the $\beta_{\phi}$ constraints, as there is no degeneracy between these two parameters. The results mainly show there is consistency in the fits to $\beta_{\phi}$ in the isotropic or anisotropic case using the pipeline of \cite{baumann2016phases}.

\subsubsection{Comparison of \emph{desilike} with \emph{Barry}}

\begin{figure}
    \centering \includegraphics[width=0.48\textwidth]{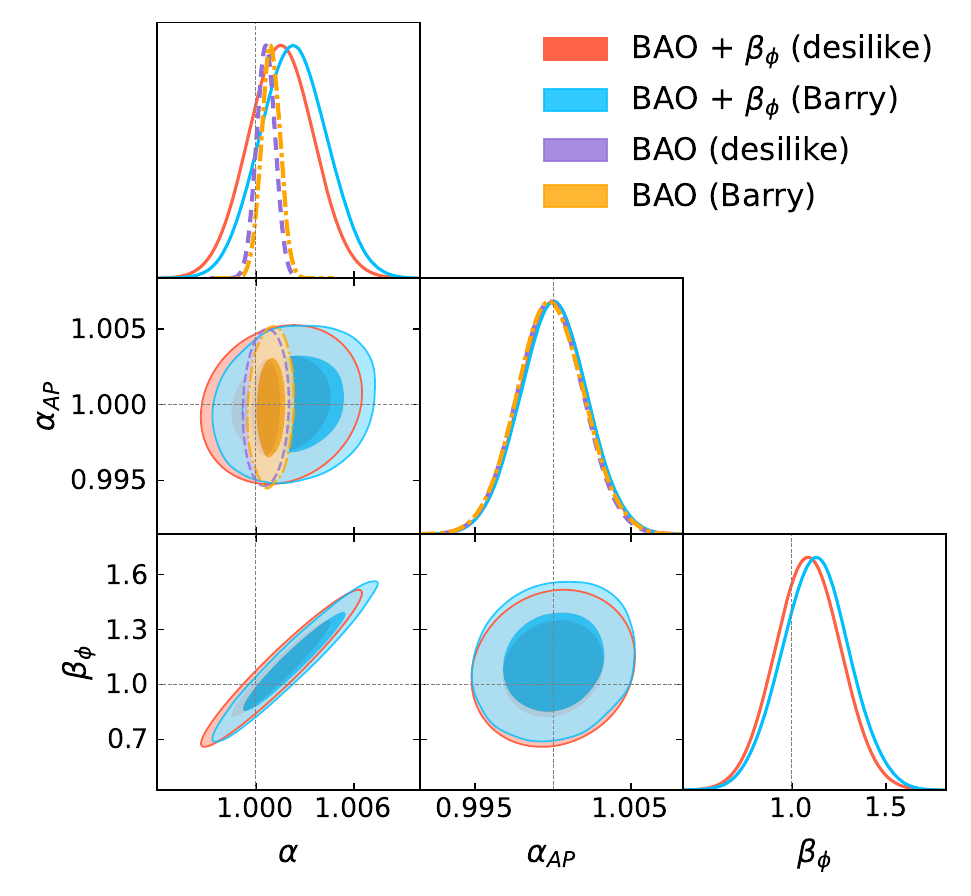}
    \caption{\small Fits to the mean of 25 CV mocks using \emph{desilike} and \emph{Barry} for the correlation function. Here we show a comparison of the results both when $\beta_{\phi}$ is fixed or allowed to vary freely. The covariance matrix has been reduced by a factor of 25 compared to a single realization. }
    \label{fig::Fig4_ELG_CV_desilikevsBarry}
\end{figure}

We additionally test the \emph{Barry} fitting code with the CV mocks to validate that the fits are consistent using different fitting tools for the anisotropic fitting pipeline. For all our fits with \emph{Barry} we use the Nautilus nested sampler \citep{nautilus}. In Figure~\ref{fig::Fig4_ELG_CV_desilikevsBarry} we show post-recon fits in comparison to fits with \emph{desilike}, where $\beta_{\phi}$ is both fixed or allowed to vary. The difference between the peak of the likelihood for $\alpha$ between the two codes when $\beta_{\phi}$ is free vs when it is fixed is $\sim 0.08\%$, and $\sim 0.03\%$ respectively. The difference in the fit to $\beta_{\phi}$ for the two codes is $\sim 4\%$, with variation occurring along the degeneracy direction for these parameters, although they are consistent within $1\sigma$. In both cases, the differences in $\alpha$ and $\alpha_{\mathrm{ap}}$ are well within the expected variance from \cite{chen2024baryon}, and we find no evidence of systematic variation between the two codes.

\begin{figure}
    \centering 
    \includegraphics[width=0.455\textwidth]{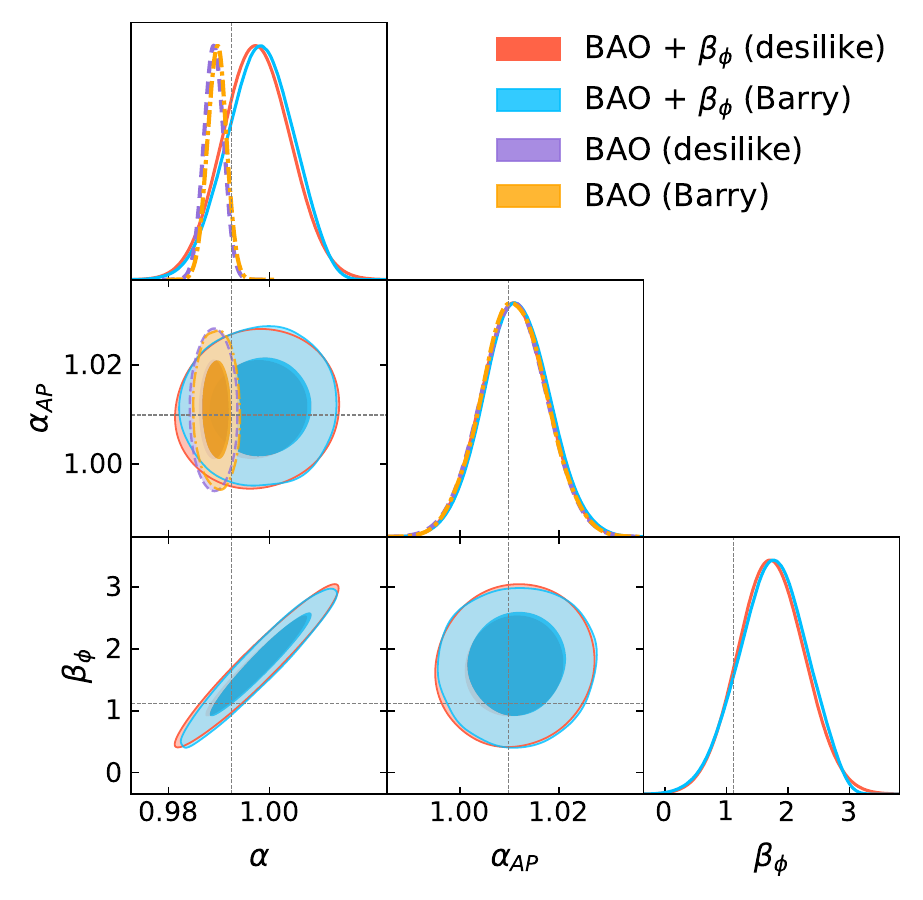}
    \caption{\small Fits to the mean of 6 power spectrum mocks using \emph{Barry} and \emph{desilike} with the polynomial broadband methodology when the true cosmology of the mocks is c003 and the fiducial and template cosmology are set to c000. This plot also allows one to compare the robustness of the fits to $\alpha$ when $\beta_{\phi}$ is or is not allowed to vary. The covariance matrix has been reduced by a factor of 6 compared to a single realization. } 
    \label{fig::comparison_c003cosmo_Pk}
\end{figure}

We also consider the robustness of the fitting methodology in the case the true cosmology of the mocks has $N_{\mathrm{eff}}$ set to a value that is not $N_{\mathrm{eff}} = 3.044$, and also in the case the fiducial or template cosmology differs from the cosmology of the mocks. Below we show the result of testing power spectrum mocks with a true underlying cosmology with $N_{\mathrm{eff}} = 3.7$, which we refer to as c003 (`CV' mocks were not available for this cosmology however). The fit to the mean of 6 mocks with a c003 cosmology is shown in Figure~\ref{fig::comparison_c003cosmo_Pk}. We use similar priors for $\Sigma_s$, $\Sigma_{\mathrm{nl, \parallel}}$, $\Sigma_{\mathrm{nl, \perp}}$ in these fits. For these mocks an analytic covariance matrix was available \citep{KP4s8-Alves}. 

We can expect when using a template with $N_{\mathrm{eff}} = 3.7$ to fit the c003 simulations we should recover $\beta_{\phi} = 1$, but for a template with $N_{\mathrm{eff}} = 3.044$ we expect that $\beta_{\phi} \approx 1.117$. The fit shown in Figure~\ref{fig::comparison_c003cosmo_Pk} is statistically consistent with this expectation, although not precise enough to differentiate clearly between the expected value and $\beta_{\phi} = 1$. However, we still see here that for these mocks there is excellent agreement between the results from our two codes, with no obvious systematic error.

\subsubsection{Robustness of modelling choices in \emph{Barry}}

\begin{figure}
    \centering \includegraphics[width=0.48\textwidth]{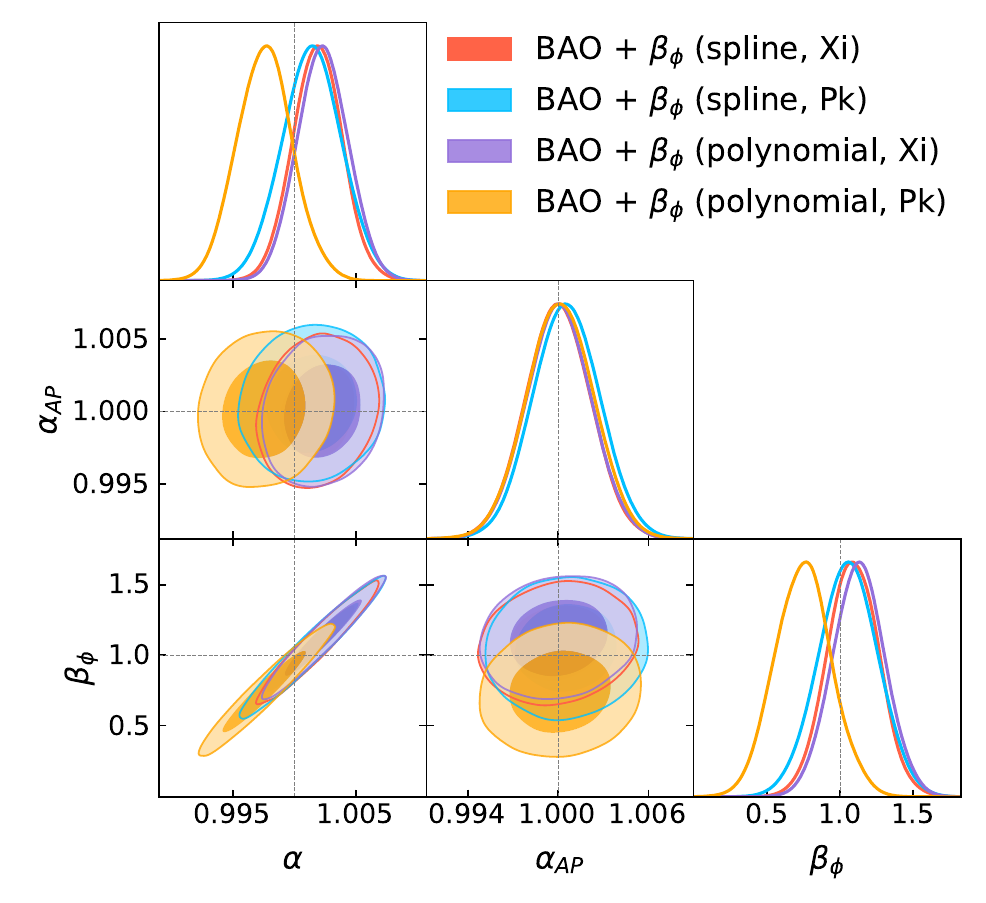}
    \caption{\small Fits to the mean of 25 CV mocks using \emph{Barry}, comparing spline and polynomial broadband methodologies. The covariance matrix has been reduced by a factor of 25 compared to a single realization.}
    \label{fig::comparison_splinepoly_PkXi}
\end{figure}

We additionally compare the results of fitting in \emph{Barry} using the polynomial broadband fitting scheme and the spline broadband fitting scheme; these broadband fitting schemes can also be accessed in \emph{desilike}. A comparison of the broadband methodologies for the fit to the mean of 25 post-recon CV power spectrum mocks and CV correlation function mocks (c000 cosmology snapshot simulations described earlier) can be seen in Figure~\ref{fig::comparison_splinepoly_PkXi}. A comparison is shown only for fits when the phase shift is allowed to vary; for fits where the phase shift is fixed (standard BAO analysis) we do not see any significant difference in the fits using either broadband methodology or choice of clustering statistic (this can be seen in Appendix~\ref{app::fitsmocksbarrybetafirstgen_spline_pkvsxi}). However, as can be appreciated from Figure~\ref{fig::comparison_splinepoly_PkXi} there is a noticeable difference in the fits to $\alpha$ and $\beta_{\phi}$ between the two different broadband methodologies for the power spectrum, although the variation to the parameters lies along their degeneracy direction and the fits are consistent with the truth within $1\sigma$.

In Table~\ref{tab::diffs_broadbandmethods_firstgenBARRY} we show the average difference between the fits to 25 mocks and the difference between the fits to the mocks means for $\beta_{\phi}$ when using the two choices of broadband fitting methodologies, for all the different cases discussed here. Considering the full set of measurements, the polynomial method applied to the power spectrum appears to be the outlier; we find a systematic difference in the fit to $\beta_{\phi}$ with a significance of $~3\sigma$ compared to results from the pipeline using the polynomial method with the correlation function or instead using the spline broadband fitting method with either the power spectrum or correlation function. We find no other significant shifts in the fit to $\beta_{\phi}$ under other comparisons between variations to the pipeline. As such, we treat the correlation function as our default going forward, interchanging between spline and polynomial broadband for various tests as these demonstrate excellent consistency. As such, we do not adopt any systematic error associated with this choice, although caution that this difference should be investigated further for future DESI data releases where we expect tighter constraints. Nonetheless, we note that (given we are fitting the average of 25 mocks) the size of this shift in $\beta_{\phi}$ (shown in Table~\ref{tab::diffs_broadbandmethods_firstgenBARRY}) is much smaller than we expect for the DESI DR1 or second-generation mock precision, and there is no clear evidence that this shift is systematic rather than statistical (given the fit still overlaps with the truth at $1\sigma$). As shown later, we also find our results for the second-generation mocks and DR1 data are robust to changes in the broadband methodology.

\begin{table}
    \centering
    \caption{\small The difference in the best fit to $\beta_{\phi}$, when different broadband methodologies are used. Here we compare the difference in the best fit to the mean of 25 mocks, $\bar{P}_{\rm{w}}$, vs the mean of separate best fits to 25 mocks for the ELGs at $z=1.1$ after reconstruction. The brackets show the significance of the difference in units of $\sigma$ (where we use the error on the mean as the uncertainty). }
    \label{tab::diffs_broadbandmethods_firstgenBARRY}
    \begin{tabular}{cccc}
        \toprule
        \midrule
         & \multicolumn{1}{c} {$|\langle\beta_{\phi,\mathrm{poly}} - \beta_{\phi,\mathrm{spline}}\rangle|_{\bar{P}_{\mathrm{w}}}$} & \multicolumn{1}{c}{$\frac{1}{25}\sum_i^{25}|\beta_{\phi,\mathrm{poly}} - \beta_{\phi,\mathrm{spline}}|_i$}  \\
        % \cmidrule(r){2-2} \cmidrule(r){3-3}
        \textbf{Mock data} & Mock mean & 25 mocks \\
        \midrule
        Pk & 0.40 (2.00) & 0.35 (1.75) \\
        Pk CV & 0.30 (2.40) & 0.29 (2.32) \\
        Xi & 0.09 (0.45) & 0.08 (0.40) \\
        Xi CV & 0.04 (0.32) & 0.05 (0.40) \\
        \bottomrule
    \end{tabular}
\end{table}

\begin{table}
    \centering
    \caption{\small A comparison of the absolute difference to the best fits to the BAO and phase shift parameters for the mean of the CV mocks, $\bar{P}_{\rm{w}}$, for ELGs at $z=1.1$, when the fitting pipeline is varied. The estimate of the significance of the difference relative to the error on the mean is given in the brackets for $\beta_{\phi}$. }
    \begin{tabular}{>{\raggedright\arraybackslash}p{2.75cm} S[table-format=1.2] S[table-format=1.4] S[table-format=1.2]}
       \toprule
       \midrule
       \textbf{Comparison} & {$100\Delta \alpha|_{\bar{P}_{\rm{w}}}$} & {$\Delta 100\alpha_{\mathrm{AP}}|_{\bar{P}_{\rm{w}}}$} & {$\Delta \beta_{\phi}|_{\bar{P}_{\rm{w}}}$}  \\ \midrule
       \emph{Barry} vs \emph{desilike} \\
       (Xi, CV, polynomial) & {0.08 (0.53)} & {0.000 (0.00)} & {0.04 (0.32)} \\[1.5mm] 
       Pk vs Xi \\
       (\emph{Barry}, CV, polynomial) & {0.47 (3.13)} & {0.010 (0.01)} & {0.38 (3.04)} \\[1.5mm]
       Pk vs Xi \\
       (\emph{Barry}, CV, spline)  & {0.05 (0.33)} & {0.070 (0.05)} & {0.04 (0.32)} \\[1.5mm]
       Spline vs polynomial \\
       (\emph{Barry}, CV, Pk) & {0.38 (2.53)} & {0.040 (0.02)} & {0.30 (1.71)} \\[1.5mm]
       Spline vs polynomial \\
       (\emph{Barry}, CV, Xi) & {0.04 (0.05)} & {0.001 (1.43)} & {0.04 (0.32)} \\[1.5mm]
       c000 vs c003 fiducial \\
       (\emph{Barry}, Pk, polynomial) & {$-$} & {$-$} & {0.18 (1.9)} \\ 
       \bottomrule
    \end{tabular}
    \label{tab::comparison_systematics_firstgenfits}
\end{table}

The effect of altering the number of free terms used in the fitting the polynomial broadband terms, the prior ranges used for the nuisance parameters, the choice of the effective redshift for the BAO template or the Boltzmann-solver used to generate the template at a fixed cosmology  \citep[which used a \textit{Planck} 2018 cosmology with $z=0.0$,][]{aghanim2020planck} and the choice of $k_{\mathrm{min}}$ used had little effect on the fits. It should be noted that there can be an impact on the best choice of priors used for nuisance parameters \emph{if} the effective redshift of the BAO template is altered significantly, because some of the nuisance parameters are redshift dependent. However, the optimal choice of priors was thoroughly tested in \cite{chen2024baryon} and we used these results to inform our choices of priors used for the fits to the mocks. To test the methodology thoroughly, we investigated the effect of altering the method used to smooth the power spectrum to isolate the wiggle component of the power spectrum  $P_{\mathrm{w}}(k)$ from the no-wiggle component $P_{\mathrm{nw}}(k)$. In \emph{Barry} there are currently three methods that can be specified to smooth the power spectrum to isolate these, however it was also found to make negligible difference to the fits of the mocks. In Table~\ref{tab::comparison_systematics_firstgenfits} we summarise the most significant differences in the fits to $\alpha$, $\alpha_{\mathrm{AP}}$ and $\beta_{\phi}$ for different modelling choices that we identified. The most significant systematic is the difference we see in the fits for the power spectrum compared to the correlation function in the case we apply fits using the polynomial broadband scheme, which further supports our choice to treat the spline-based method as our default. 

\begin{figure}
    \centering \includegraphics[width=0.48\textwidth]{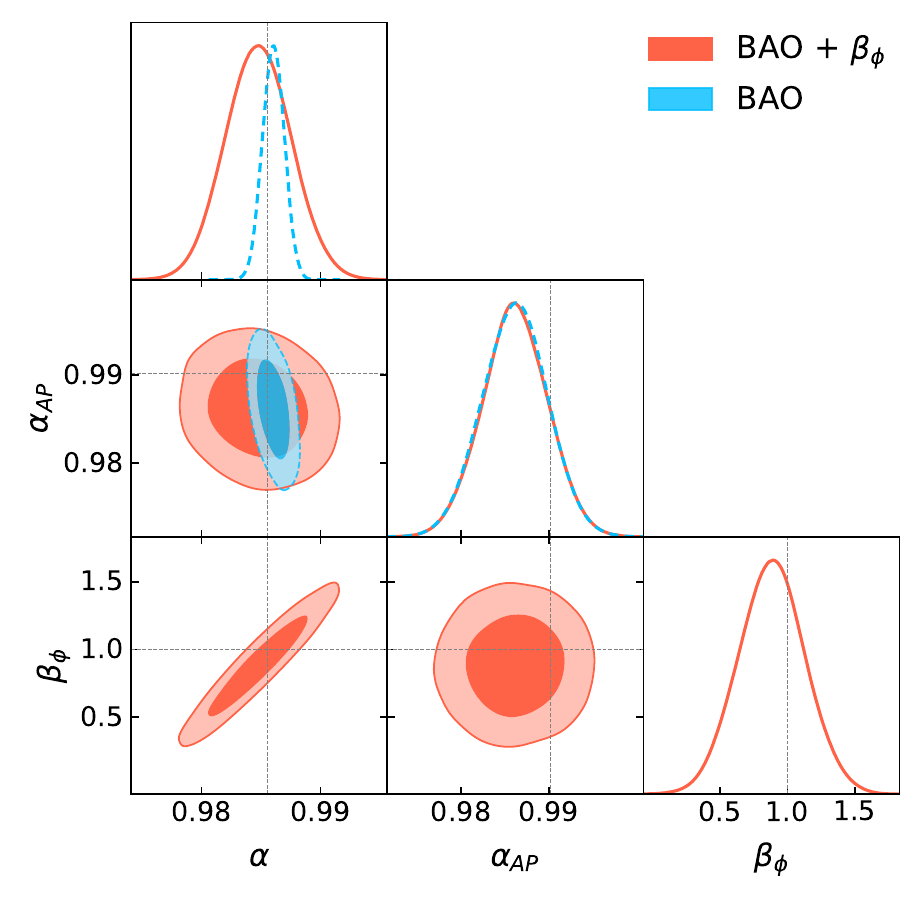}
    \caption{\small Similar to Figure~\ref{fig::comparison_c003cosmo_Pk}, however only showing the fit to the mean of 25 CV mocks with the polynomial broadband methodology in the case where the template and true cosmology are both set to c000, while the fiducial cosmology is set to c003. In this case, we still expect to recover $\beta_{\phi} = 1.0$. These fits demonstrate that our method is robust to the choice of fiducial cosmology.} \label{fig::comparison_c000cosmo_Pk_0003fiducial}
\end{figure}

It is interesting to test whether we still recover $\beta_{\phi} = 1$ in the case we alter the fiducial cosmology. This is because in the case the template is set to have $N_{\mathrm{eff}} = 3.044$ such that it is the same as the true cosmology, we expect $\beta_{\phi} = 1$, regardless of the choice of fiducial cosmology for the clustering (this does not apply for $\alpha$ and $\alpha_{\mathrm{AP}}$, whose truth depends on both the true, fiducial and template cosmology). As mentioned earlier, this is because changing $N_{\mathrm{eff}}$ (and thus $\beta_{\phi}$) in the fiducial cosmology should not alter the fit to $\beta_{\phi}$ since $\beta_{\phi}$ arises as a physical shifting of the $k$ modes in the BAO wiggles rather than a geometric distortion (AP Effect) that arises from transforming between redshift-space and real-space co-ordinates. To test this, we fit the mean of the 25 mocks (non CV mocks) again in Figure~\ref{fig::comparison_c000cosmo_Pk_0003fiducial}. We use c000 mocks (for the truth and template) where the fiducial cosmology is set to the c003. For brevity we show only the results using the polynomial broadband methodology and we show the absolute difference in the fit to $\beta_{\phi}$ in this scenario in Table~\ref{tab::comparison_systematics_firstgenfits}. Note this \emph{differs} from Figure~\ref{fig::comparison_c003cosmo_Pk}, because here we have altered the fiducial cosmology rather than altering the real cosmology. The effect of the fiducial (grid) cosmology in DESI mocks for the standard BAO analysis is studied in detail in \cite{perez2024fiducial} for the more sophisticated `second-generation' mocks. 

\subsubsection{Summary}

Overall, our results show a clear degeneracy between $\alpha$ and $\beta_{\phi}$, which indicates it may be important to include $\beta_{\phi}$ as a free parameter when fitting the BAOs generally (although as we will show later, applying an informative prior to this recovers the published DESI Year 1 results \citealt{adame2024desi3}). Of course, when comparing the systematic differences to the fits in $\alpha$ for the standard case vs the case where we include $\beta_{\phi}$, the impact is insignificant, due to the fact $\beta_{\phi}$ is weakly constrained. However, for future data we may expect it to be possible the degeneracy between $\alpha$ and $\beta_{\phi}$ could lead to small systematic biases in the case $\beta_{\phi}$ is fixed to one but the data prefers a different value.%

From Table~\ref{tab::comparison_systematics_firstgenfits} and Figure~\ref{fig::comparison_splinepoly_PkXi}, it is apparent that the fits using the polynomial broadband method for the power spectrum data results in a difference between polynomial and spline fitting methodologies at a significance of $3\sigma$; for fits where $\beta_{\phi}$ is a free parameter, $\sigma_{\mathrm{sys}}(\beta_{\phi}) = \pm 0.38$. Given we find no other methodologies result in such systematics, we avoid using the polynomial fitting method for power spectrum fits for the second-generation mocks and data. As mentioned previously, we find all other choices explored for the fitting methodology do not show significant evidence of systematic differences between the best fits to the BAO parameters or $\beta_{\phi}$ relative to the statistical uncertainty in the fits, although we caution this will need to be considered carefully for future DESI data releases. 

\subsection{Fitting DESI second-generation mocks with \emph{desilike}}\label{sec::fits_secondgenmocks}

\begin{figure*}
    \begin{subfigure}[b]{1\textwidth} 
    \centering \includegraphics[width=1\textwidth]{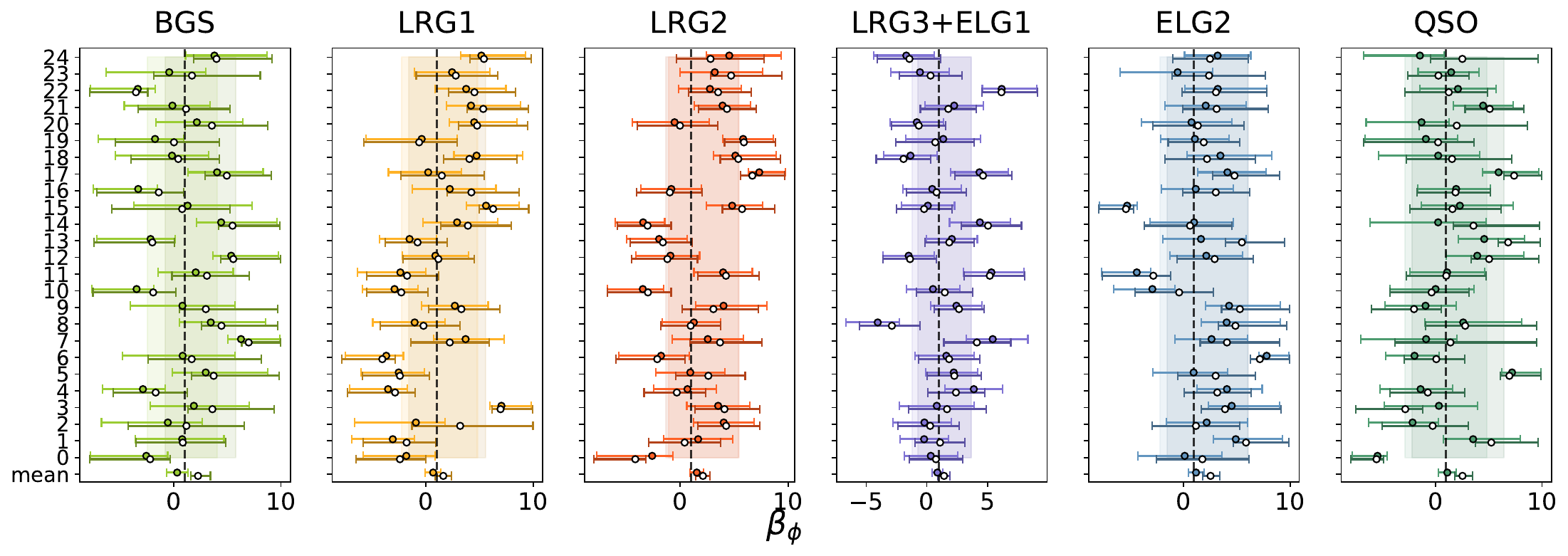}
    \end{subfigure} 
    \begin{subfigure}[b]{1\textwidth} 
    \centering \includegraphics[width=1\textwidth]{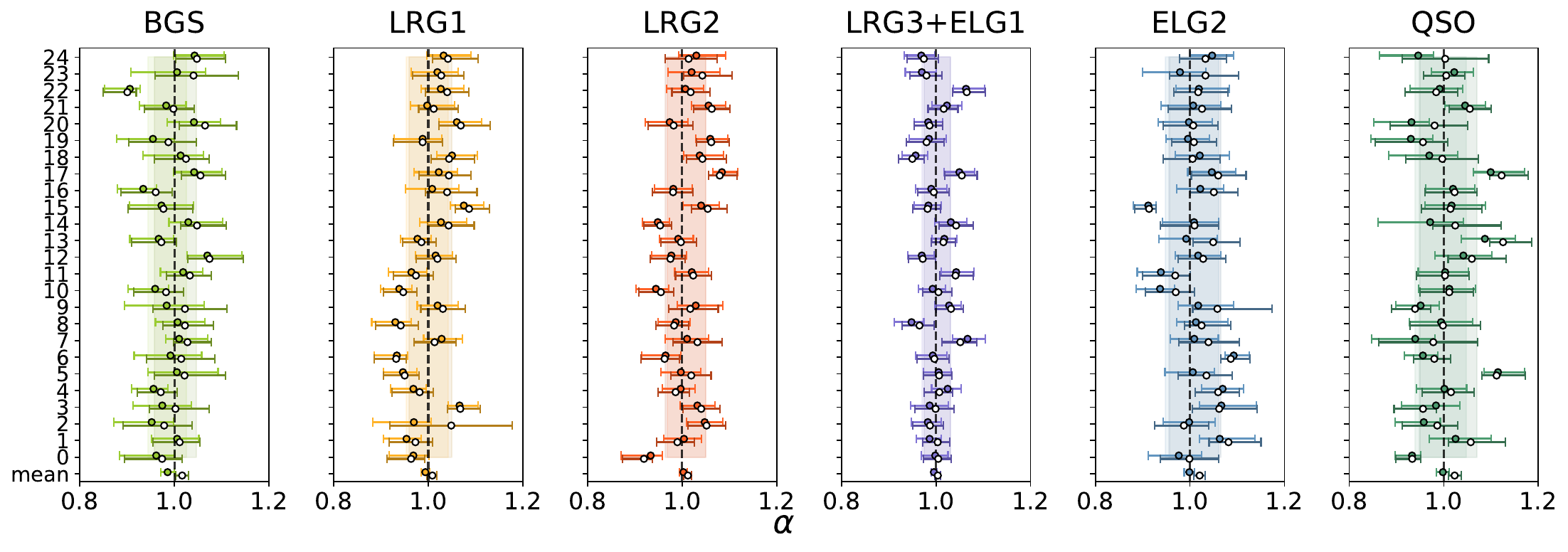}
    \end{subfigure} 
    \begin{subfigure}[b]{1\textwidth} 
    \centering \includegraphics[width=1\textwidth]{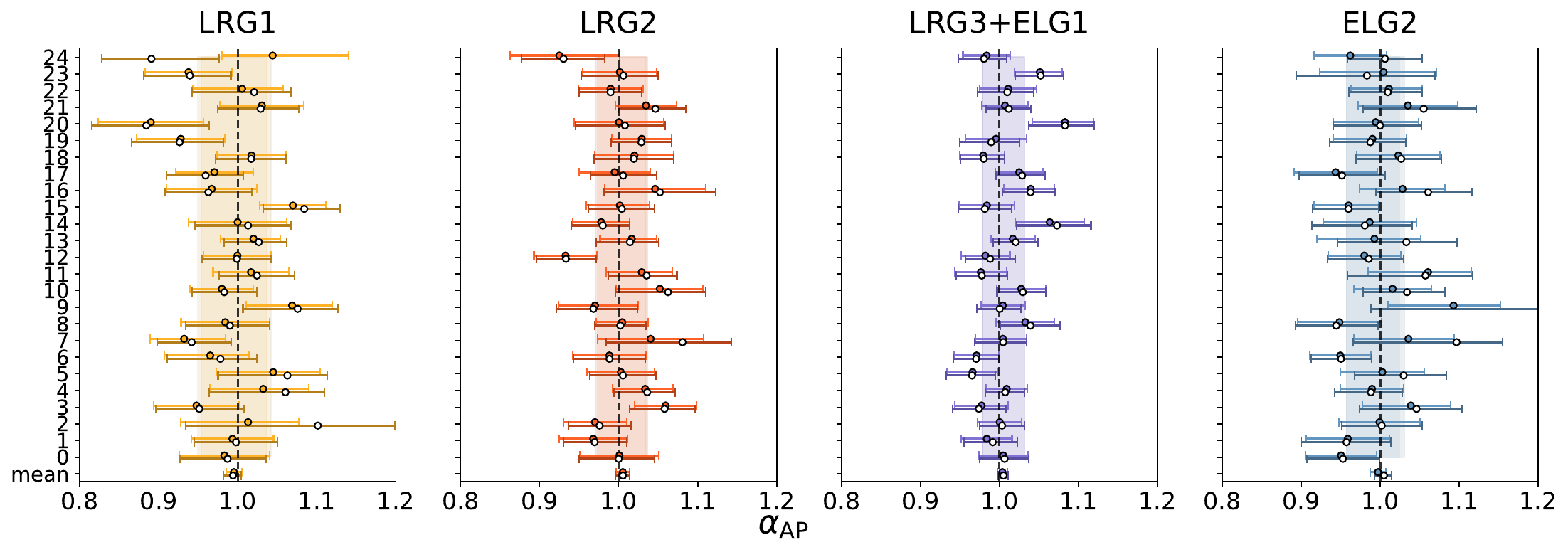}
    \end{subfigure} 
    \caption{Our fits of the 25 mocks and the mean for $\beta_{\phi}$ (top panel), $\alpha$ (middle panel) and $\alpha_{\rm{AP}}$ (lower panel), for the 6 DESI samples considered here. The coloured points show \textit{desilike} fits while the white points show the fits using \textit{Barry}. Shaded regions show one standard deviation about the mean fit for the 25 mocks, where the darker colored shaded regions are \textit{desilike} and the lighter are \textit{Barry}. The mean and standard deviation has been weighted by the uncertainty on each mock, so that more poorly constrained fits have less weight in the mean. For the BGS and QSO samples, we only fit an isotropic BAO model, and hence do not report $\alpha_{\mathrm{AP}}$.}
\label{fig::Fig8_7tracersX25mocksandMean_secondgen}
\end{figure*}

\begin{table*}
    \centering
    \caption{Average of the difference in the fits to $\alpha$, $\alpha_{\mathrm{AP}}$ and $\beta_{\phi}$ between \textit{Barry} and \textit{desilike}. The numbers in brackets also give this uncertainty in the number of standard deviations (taking the weighted standard deviation of the 25 mocks) - thus quantifying the significance of the difference in numbers of $\sigma$. We also show for comparison the difference between the fits to the mock means, $\bar{P}_{\rm{w}}$, for each parameter. }
    \label{tab::comparison-secondgen}
    \begin{tabular}{>{\raggedright\arraybackslash}p{1.5cm}ccc|ccc}
        \toprule
        \midrule
        & \multicolumn{3}{c}{$\Delta x$ from the weighted mean of 25 mocks, $\frac{1}{25} \sum_{i}^{25} | x_{\mathrm{Barry}} - x_{\mathrm{desilike}} |$ ($\frac{\Delta x}{\sigma}$)} & \multicolumn{3}{c}{$\Delta$ from fit to data mock mean $\bar{P}_{\rm{w}}$} \\
        \textbf{Tracer} & {$100 \Delta \alpha$} & {$100 \Delta \alpha_{\mathrm{AP}}$} & {$\Delta \beta_{\phi}$} & {$100 \Delta \alpha|_{\bar{P}_{\rm{w}}}$} & {$100 \Delta \alpha_{\mathrm{AP}}|_{\bar{P}_{\rm{w}}}$} & {$\Delta \beta_{\phi}|_{\bar{P}_{\rm{w}}}$} \\
        \midrule
        %BGS & 1.7 (2.06) & {$-$} & 1.549 (2.34) & 2.9 & {$-$} & 1.782 \\ poly

        BGS & 1.7 (2.1) & {$-$} & 1.711 (2.62) & 0.03 & {$-$} & 1.950 \\  %spline 
        
        %LRG1 & 0.6 (0.65) & 0.1 (0.11) & 0.557 (0.77) & 1.5 & 0.3 & 0.966 \\ %poly

        LRG1 & 0.7 (0.8) & ${<0.1}$ ${(< 0.1)}$ & 0.570 (0.80) & 1.5 & 0.1 & 0.947 \\ %spline
        
        %LRG2 & 0.2 (0.20) & {$<0.1$} (0.05) & 0.124 (0.99) & 1.0 & 0.3 & 0.625 \\ %poly

        LRG2 & 0.2 (0.2) & ${<0.1}$ ${(< 0.1)}$ & 0.550 (0.82) & 0.9 & ${< 0.1}$ & 0.570 \\ %spline  

        %LRG3+ELG1 & 0.2 (0.36) & {$< 0.01$} (0.02) & 0.242 (0.45) &  0.9 & 0.2 & 0.551 \\ %poly

        LRG3+ELG1 & 0.3 (0.5) & ${<0.1}$ ${(< 0.1)}$ & 0.239 (0.49) & 0.8 & 0.1 & 0.550 \\ %spline
        
        %ELG2 & 1.5 (1.32) & 0.5 (0.93) & 0.879 (0.10) & 2.2 & 0.7 & 1.357 \\  %poly

        ELG2 & 0.7 (0.6) & 0.3 (0.5) & 0.361 (0.44) & 2.2 & 0.7 & 1.361 \\  %spline
        
        %QSO & 1.6 (1.54) & {$-$} & 0.886 (1.18) & 2.8 & {$-$} & 1.592 \\ %poly

        QSO & 1.6 (1.5) & {$-$} & 1.13 (1.47) & 2.5 & {$-$} & 1.425 \\ %spline

        \bottomrule
    \end{tabular}
\end{table*}

\begin{table}
    \centering
    \caption{Summary of the choices of priors and ranges used for fitting the second-generation mocks and DR1 data for each DESI tracer. The subscript $i$ denotes different tracer bins (BGS, LRG1, LRG2 etc.) and thus allows us to specify cases where the prior differs for specific cases which are otherwise the same. }
    \label{tab::propertiespriors-secondgen}
    \begin{tabular}{>{\raggedright\arraybackslash}p{1.7cm} ll p{1.7cm}}
        \toprule
        \midrule
         & \multicolumn{2}{c}{Gaussian prior: $\mathcal{N}(\mu, \sigma)$} \\
        \cmidrule(r){2-3}
        \textbf{Parameter} & Post-recon & Pre-recon &  \\
        \midrule
        {$\Sigma_S$} ($\mathrm{Mpc} h^{-1}$) & $\mathcal{N}(2,2)$ & $\mathcal{N}(2,2)$ \\{$\Sigma_{\mathrm{nl,\parallel}}$} ($\mathrm{Mpc} h^{-1}$) & $\mathcal{N}(6,2)$ & $\mathcal{N}(x_i,2)$ & $x_{\rm{ELG2}}=8.5$ else $x_i = 9$  \\ 
        {$\Sigma_{\mathrm{nl}}$} ($\mathrm{Mpc} h^{-1}$) & $\mathcal{N}(x_i,2)$ & $\mathcal{N}(y_i,2)$ & $x_{\rm{BGS}} = 8$, $x_{\rm{QSO}} = 6$, $y_{\rm{BGS}} = 10$, $y_{\rm{QSO}} = 9$\\ 
        
        {$\Sigma_{\mathrm{nl,\perp}}$} ($\mathrm{Mpc} h^{-1}$) & $\mathcal{N}(3,1)$ & $\mathcal{N}(x_i,1)$ & $x_{\rm{BGS}} = 6.5$, $x_{\rm{QSO}} = 3.5$ else $x_i = 4.5$ \\ 
        
        {$b$} & $\mathcal{U}(0.5, 4)$ & $\mathcal{U}(0.5, 4)$  \\ 
        %{$\beta=b/f$} & NA & NA & [0.1, 0.7] \\ 
        {$\beta_{\phi}$} & $\mathcal{U}(-8, 10)$ & $\mathcal{U}(-8, 10)$ \\ 
        {$\alpha$} & $\mathcal{U}(0.8, 1.2)$ & $\mathcal{U}(0.8, 1.2)$  \\ 
        {$\epsilon = (\alpha_{\mathrm{AP}}^{1/3} - 1)$} (\emph{Barry}) & $\mathcal{U}(-0.2, 0.2)$ & $\mathcal{U}(-0.2, 0.2)$ \\ 
        {$\alpha_{\mathrm{AP}}$} (\emph{desilike}) & $\mathcal{U}(0.8, 1.2)$ & $\mathcal{U}(0.8, 1.2)$ \\ 
        \bottomrule
    \end{tabular}
\end{table}

In this section we show the results of fitting realistic second-generation mocks. Unlike the first-generation mocks, these include potential data systematics and selection functions matching the DESI data \citep{adame2024desi3, adame2024desi}. We explore fits in which we do not reduce the covariance on the mocks in order to see the expected constraining power for $\beta_{\phi}$ with the realistic uncertainty for a single realisation. For these fits, the covariance matrices for the correlation function are semi-analytic and are created using \textit{RascalC} \citep{philcox2020rascalc, rashkovetskyi2024semi},\footnote{\url{https://github.com/misharash/RascalC-scripts/tree/DESI2024/DESI/Y1}} while for the power spectrum the covariances are made using \textit{covaPT} \citep{wadekar2020galaxy, KP4s8-Alves}.\footnote{\url{https://github.com/cosmodesi/thecov}}

\subsubsection{Fits to \texorpdfstring{$\beta_{\phi}$}{TEXT} from individual tracers}\label{sec::fitssecondgenindividual}

Figure~\ref{fig::Fig8_7tracersX25mocksandMean_secondgen} shows fits to the individual mocks for the correlation function of each tracer produced using \emph{desilike} and \emph{Barry}; we show fits with the post-recon mocks. We only show the fits using the spline broadband methodology. 
%, in addition to choosing to fit the post-recon correlation function. 
As with the real data results in section~\ref{sec::resultswithyr1data}, our `baseline' results will use the `spline' methodology to be consistent with \cite{adame2024desi, adame2024desi3}. However, we expect based on the analysis of systematics from first-generation mocks presented earlier, the second-generation fits will be independent of the choice between a spline or polynomial broadband approach as they are consistent when using the correlation function. For ELG1 and LRG3 which overlap in redshift, we use a combined mock for the correlation function (the same is used for the fits to the real data). The priors we use on nuisance parameters for second-generation mocks are specified in Table~\ref{tab::propertiespriors-secondgen}. For polynomial fits we set the coefficients for the powers in the fit the broadband shape as $[-2, -1, 0]$ and $[-1, 0, 1, 2, 3]$ for the correlation function and power spectrum respectively. Seven coefficients are used to fit the broadband shape using the spline methodology for power spectrum fits (and as before, the Hankel transformation of the cubic splines to configuration space results in fitting just two coefficients, in addition to two more coefficients for polynomial terms).

It is clear that there is consistency in the fits between \emph{Barry} and \emph{desilike} tracers. The only significant differences are for fits in cases where neither \emph{Barry} or \emph{desilike} obtains a strong constraint on the BAO parameters and $\beta_{\phi}$. In such cases, $\beta_{\phi}$ can hit the end range of the priors used, and we may expect some differences due to differences in how the sampling is performed or due to, subtle differences in the modelling of the Kaiser factor between \textit{desilike} and \textit{Barry} (see discussion in \citealt{adame2024desi3}, although it is expected that these model differences will have negligible effect on the fits to the BAO parameters and $\beta_{\phi}$ and only affect the fits to the nuisance parameters involving the galaxy bias). For the more constraining tracers (i.e., LRG3+ELG1), we see excellent agreement between the codes and both \emph{Barry} and \emph{desilike} obtain unbiased constraints on the phase shift, which validates our methodology for application to the DR1 data. Table~\ref{tab::comparison-secondgen} shows the difference (on average) between the fit to each BAO parameter + $\beta_{\phi}$ between \textit{Barry} and \textit{desilike}, and for comparison the difference between the fit to the mock means. This, like previously for the first-generation mocks, allows us to verify any potential systematic uncertainties on the fits to the parameters due to differences between the fitting codes, which can be considered in the error budget for the fits to the data. However, for each of the tracers, the relative difference in the fits to each parameter to the statistical uncertainty on the mean is $\lesssim2.1\sigma$ for $\alpha$, $\lesssim 2.6\sigma$ for $\beta_{\phi}$ and $\lesssim 0.5\sigma$ for $\alpha_{\mathrm{AP}}$; the more weakly constraining tracers tend to show larger differences. This indicates no strong detections of any systematic differences between the codes, and that any systematics are far below the expected precision of the Year 1 data (noting that the statistical error on the data is expected to be 5 times larger than the error on the mean used here).

\subsubsection{Determining a combined fit to \texorpdfstring{$\beta_{\phi}$}{TEXT} from multiple tracers}\label{sec::combinedfits}

\begin{figure}
    \centering
    \includegraphics[width=0.4\textwidth]{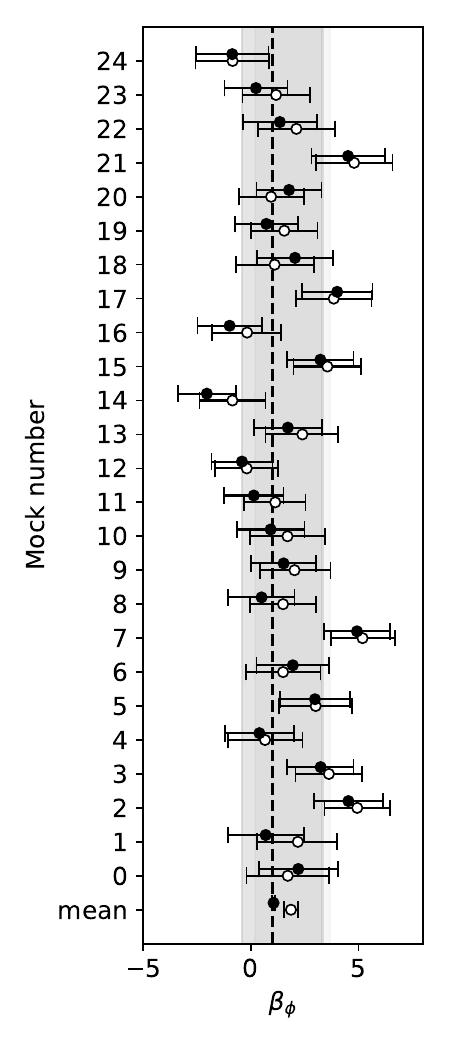}
    \caption{Our fits to $\beta_{\phi}$ from the combinations of the LRGs and ELGs (all tracer bins except QSOs and BGS) with the second-generation mocks, for \textit{desilike} (black points) and \textit{Barry} (white points). The shaded regions show the (weighted) 1-standard deviation about the weighted mean for the fits to individual mocks, the fit to the mean of the mocks is shown at the bottom of the plot.}
    \label{fig::combined_fits_on_mocks}
\end{figure}

We are also interested in looking at the combined constraining power for $\beta_{\phi}$ from the tracers in addition to the constraints from individual tracers. In order to do this, we take a mock for each tracer, then importance sample the individual constraints LRG1, LRG2 and ELG2 to weight the fit to $\beta_{\phi}$ for the tracer that obtains the most constraining fits, LRG3+ELG1. In this process, we are effectively computing the posterior distribution of $\beta_{\phi}$ from a combined fit of LRG1, LRG2, LRG3+ELG1 and ELG2,
\begin{equation}
    P_{\mathrm{combined}}(\beta_{\phi}) \propto \prod_{i} \int d\alpha_i\,d\alpha_{\mathrm{AP},i}\,P_{i}(\alpha_i, \alpha_{\mathrm{AP},i}, \beta_{\phi}|\xi_{i}),
    \label{eq:combined}
\end{equation}
where the index $i$ runs over LRG1, LRG2, LRG3+ELG1 and ELG2 and the individual $\alpha$ and $\alpha_{\mathrm{AP}}$ constraints are treated as independent, while $\beta_{\phi}$ is shared. Given that BGS and QSOs obtain weak constraints on $\beta_{\phi}$, we only use LRGs and ELGs for the combined fits to the mocks and data. This result, which we will refer to as `combined fits' to the mocks can be seen in Figure~\ref{fig::combined_fits_on_mocks}. This shows we can obtain reasonable fits to $\beta_{\phi}$ using this methodology and find that combining the multiple tracers improves the uncertainty on $\beta_{\phi}$ by a factor $\sim 1.9$ compared to the fits to LRG3+ELG1 alone. The fits are also consistent between \emph{Barry} and \emph{desilike} when combining the tracers in this way. The fits here are done using individual mock fits with the `baseline' methodology. The average absolute difference of each mock fit to the combination of the tracers, for fits to the tracers done using \textit{desilike} and \textit{Barry} is $\Delta \beta_{\phi} = 0.431$ (or $\frac{\Delta}{\sigma} = 1.282$ where $\sigma$ is the weighted standard error of the mean of the combined fits). The absolute difference
between the fits to the mock mean data (average correlation function of 25 mocks) is $\Delta \beta_{\phi} = 0.809$. 

\begin{figure*}
    \centering 
    \begin{subfigure}[b]{1\textwidth} 
    \includegraphics[width=1.0\textwidth]{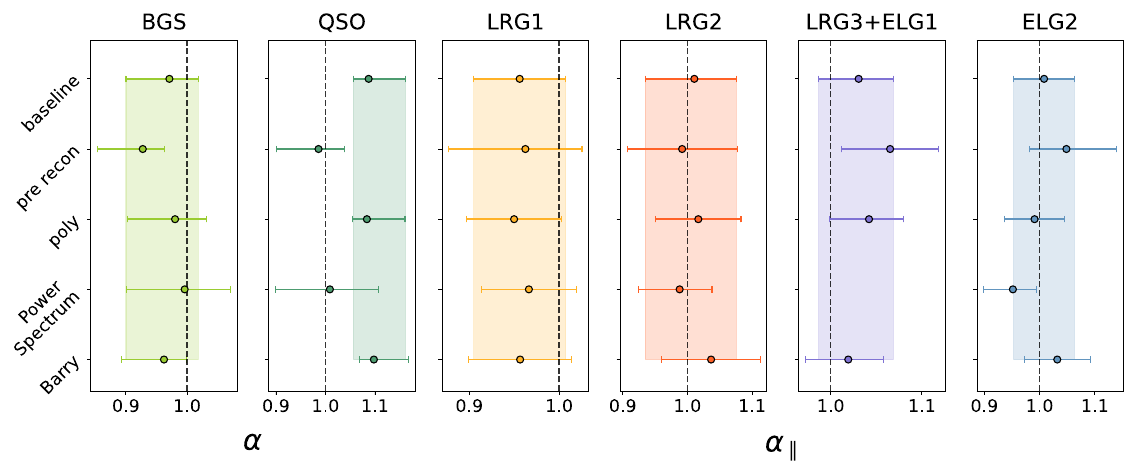}
    \end{subfigure} 
    \begin{subfigure}[b]{1\textwidth} 
    \includegraphics[width=1\textwidth]{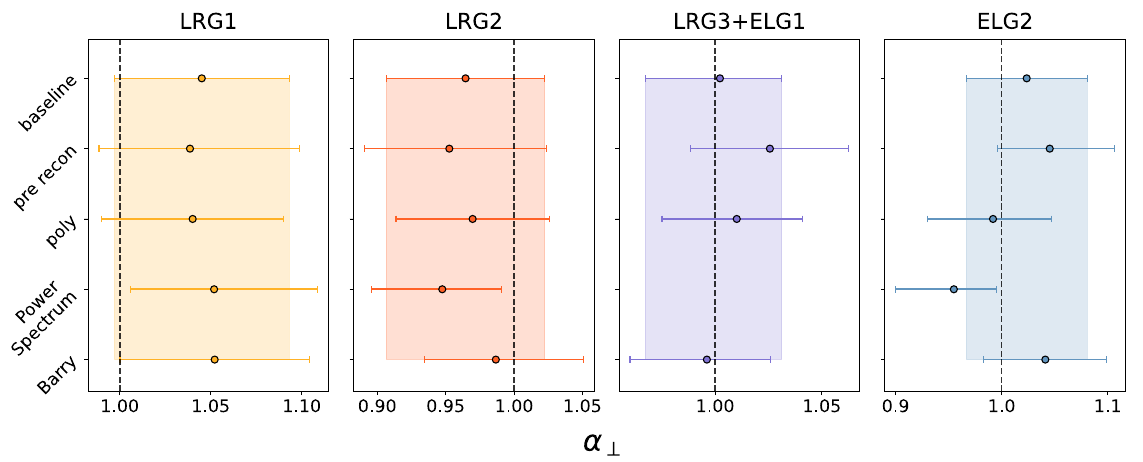}
    \end{subfigure} 
    \begin{subfigure}[b]{1\textwidth} 
    \includegraphics[width=1\textwidth]{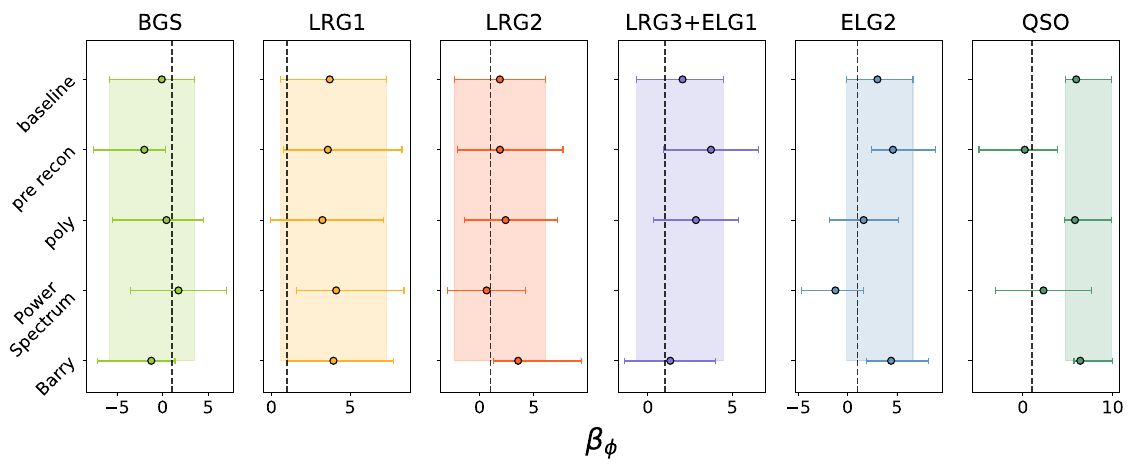}
    \end{subfigure} 
    \caption{Our fits to the DESI tracers in DR1. For ease of interpretation for the distortions to physical distances, we plot $\alpha_{\parallel}$, $\alpha_{\perp}$ in place of $\alpha$ and $\alpha_{AP}$ and additionally $\beta_{\phi}$. For BGS and QSOs we only study the isotropic scaling; as such we only plot $\alpha$. } \label{fig::results_YR1_variousanalysischoicesfits_tracers}
\end{figure*}

\begin{figure*}
    \centering \includegraphics[width=1.0\textwidth]{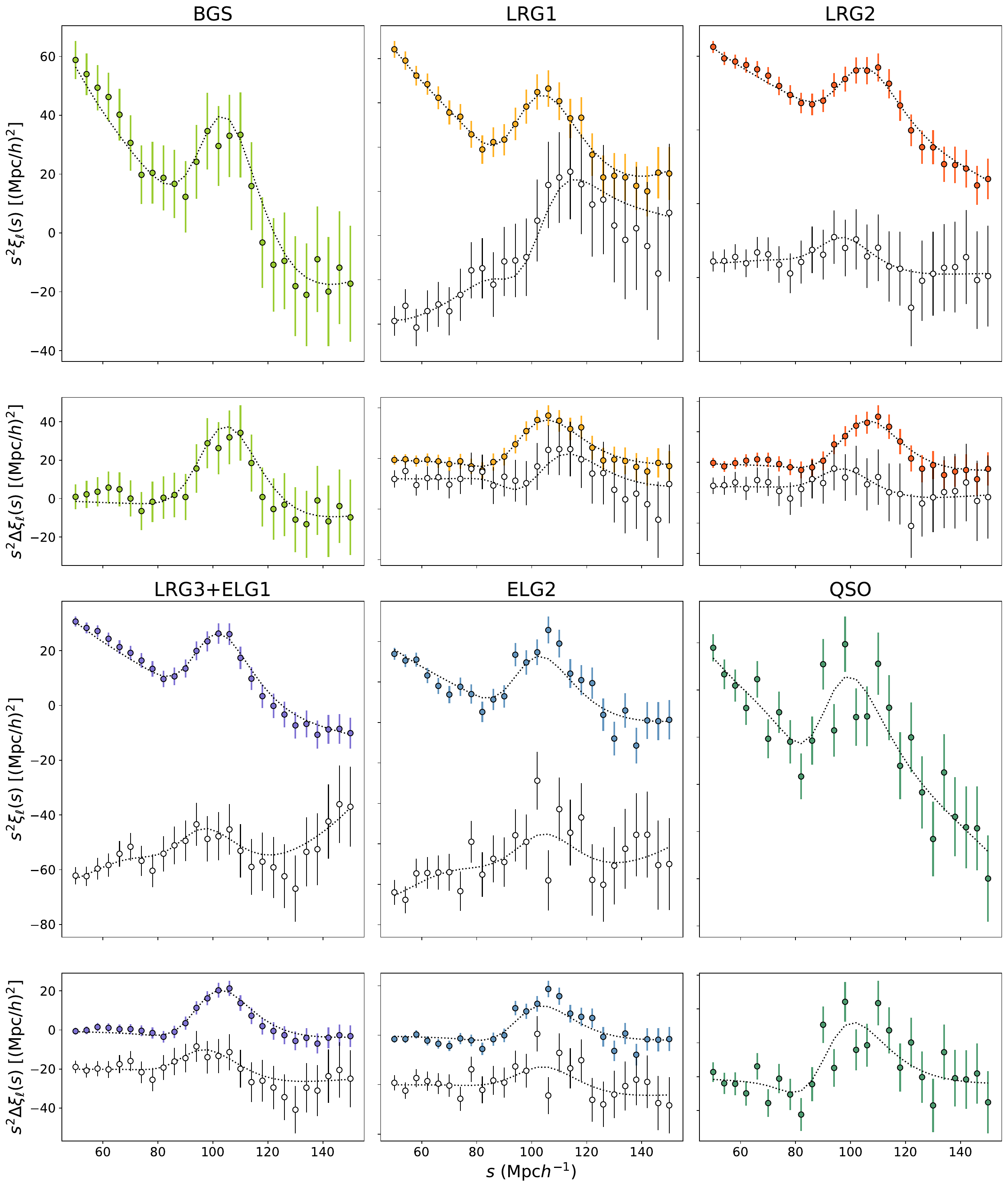}
    \caption{The best fit models for our the DESI tracers in the DR1 data with $\beta_{\phi}$ varying in the analysis, for comparison to the data. The coloured data shows the best fit models and data for the monopole, and the grey shows the quadrupole. The smaller panels show the isolated BAO peak for each tracer by subtracting the smooth model from the data. For visual clarity a small vertical offset has also been applied to the quadrupole for these panels.} \label{fig::results_YR1_datavsmodels_tracers_baseline}
\end{figure*}

\section{Results with DR1 DESI data}\label{sec::resultswithyr1data}

Having studied the robustness of our fitting methodology to different BAO fitting software and analysis choices in the previous section, we present fits to the BAO parameters and phase shift $\beta_{\phi}$ with the DR1 data. In all cases, we use the same choices of coefficients for the broadband fitting and choices of priors as done for our fits to second-generation mocks. These choices are the same as those used for the BAO results presented by \cite{adame2024desi, adame2024desi3}. For the remainder of this paper, unless stated otherwise, we will refer to and present our main results using the `baseline' methodology; as before, the `baseline' fits involve post-recon fits to the correlation function using the spline fitting methodology and using the \emph{desilike} software. 

\subsection*{Constraints from individual tracers}

In Figure~\ref{fig::results_YR1_variousanalysischoicesfits_tracers} we show the individual fits to the BAO distortion parameters $\alpha_{\parallel}$ and $\alpha_{\perp}$ and phase shift $\beta_{\phi}$ for tracers for the DR1 data. The figure also allows a comparison of the baseline results to various individual changes to the fitting pipeline, allowing one to see how robust our results are. One can see how the best fits vary for each tracer under various choices;
\begin{itemize}
    \item when we do not apply reconstruction to the data,  
    \item when we use the polynomial broadband fitting methodology rather than the spline approach, 
    \item when we fit the power spectrum rather than the correlation function,
    \item or when we produce the fits with \emph{Barry} rather than \emph{desilike}.
\end{itemize}
Compared to the fits presented in \cite{adame2024desi3} for DESI DR1 (see Figure 12 in their work which is analogous to our Figure~\ref{fig::results_YR1_variousanalysischoicesfits_tracers}), there is slightly less consistency in our fits presented when the pipeline is varied; the results shown in \cite{adame2024desi3} are more statistically robust to changes. This is most noticable (and expected) for the pre-recon fits, due to the fact that not applying reconstruction significantly weakens the constraints. However, we can expect that for all of the fits, the strong degeneracy between $\alpha$ and $\beta_{\phi}$ should lead to less constraints that move along this degeneracy direction due to statistical noise. We also do not expect very strong constraints from BGS or QSO tracers given the results with the mocks showed that these tracers are much less constraining when $\beta_{\phi}$ is allowed to vary, and this is apparent in our results. 

It is also important to validate that our fits to $\alpha$ and $\alpha_{\rm{AP}}$ are consistent with those of the DESI DR1 results presented in \cite{adame2024desi, adame2024desi3} and that the additional uncertainty in our fits when $\beta_{\phi}$ is allowed to vary does not indicate that the uncertainty in the DESI DR1 results are underestimated. To check this, we recalculated these parameters in a thin slice of the parameter space for $\beta_{\phi}$ by importance sampling the data to impose a prior. This was done by taking the \textit{Planck} 2018 MCMC chain for $N_{\rm{eff}}$ from fits to the $\Lambda$CDM + $N_{\rm{eff}}$ model, which gives $N_{\rm{eff}} = 2.99 \pm 0.17$. In principle, we expect that this is equivalent to rerunning the fits with a very tight informative prior on $\beta_{\phi}$ from \textit{Planck}. In doing so, we found the best fits to the parameters expressed as $\alpha$ and $\alpha_{\mathrm{AP}}$ were fully consistent with the DR1 results presented in \cite{adame2024desi3}. The fits and comparison to the numbers in \cite{adame2024desi3} can be seen in Table~\ref{tab::results_tightpriorbeta}. The $1\sigma$ uncertainty on $\alpha_{\parallel}$ varies by approximately $5-8$\% for different tracers, and for $\alpha_{\perp}$ there is a $3-15$\% variation in the $1\sigma$ uncertainty. We additionally looked at the constraints in the case when the width of the \textit{Planck} prior for $N_{\mathrm{eff}}$ was increased by a factor of 10 times; the fits are still consistent with those presented in \cite{adame2024desi3} and only the fits to $\beta_{\phi}$ are weakened. 

\begin{table}
    \centering
    \caption{\small Constraints on $\beta_{\phi}$, $\alpha$ and $\alpha_{\rm{AP}}$ from the DR1 data baseline results. We also include the case where a tight prior is included on $\beta_{\phi}$ by importance sampling with the \textit{Planck} 2018 $\Lambda$CDM + $N_{\mathrm{eff}}$ MCMC chain. Results quoted from \protect\cite{adame2024desi3} are also presented for comparison, demonstrating excellent agreement between our work at previous DESI results when an informative prior on $\beta_{\phi}$ is included.}
    \label{tab::results_tightpriorbeta}
    \begin{tabular}{p{1.7cm} S[table-format=15] S[table-format=15.5] S[table-format=15.5] S[table-format=15.5]}
        \toprule
        \midrule
        \textbf{Data} & $\beta_{\phi}$ & $\alpha$ & $\alpha_{\rm{AP}}$ \\
        \midrule
        \multicolumn{4}{l}{($\beta_{\phi}$ free)} \\
        \midrule
        BGS & {$-0.1^{+4.2}_{-5.0}$} & {$0.971^{+0.05}_{-0.066}$} & {$-$} \\[1mm]
        LRG1 & {$3.7\pm 3.2$} & {$1.014\pm 0.044$} & {$0.915\pm 0.043$}\\[1mm] 
        LRG2  &  {$1.9\pm 4.2$} & {$0.979\pm 0.057$} & {$1.048\pm 0.048$}\\[1mm] 
        LRG3+ELG1 & {$2.1\pm 2.6$} & {$1.012^{+0.031}_{-0.035}$} & {$1.028\pm 0.030$}\\[1mm] 
        ELG2  & {$3.0^{+3.6}_{-3.2}$} & {$1.018\pm 0.052$} & {$0.986\pm 0.045$}\\[1mm] 
        QSO & {$5.9^{+4.0}_{-2.3}$} & {$1.087^{+0.073}_{-0.045}$} & {$-$} \\[1mm]
        \midrule
        \multicolumn{3}{l}{($\beta_{\phi}$ free + \textit{Planck} prior on $N_{\rm{eff}}$)} \\ \midrule
        BGS & {$0.97\pm 0.04$} & {$0.982\pm 0.020$} & {$-$} \\[1mm]
        LRG1  & {$0.97\pm 0.04$} & {$0.979\pm 0.012$} & {$0.915\pm 0.041$}\\[1mm] 
        LRG2  & {$0.98\pm 0.04$} & {$0.966\pm 0.013$} & {$1.047^{+0.047}_{-0.052}$}\\[1mm] 
        LRG3+ELG1 & {$0.97\pm 0.04$} & {$0.998\pm 0.008$} & {$1.026\pm 0.029$}\\[1mm] 
        ELG2  & {$0.97\pm 0.04$} & {$0.987\pm 0.015$} & {$0.988\pm 0.043$}\\[1mm] 
        QSO & {$0.98\pm0.04$} & {$0.998\pm 0.024$} & {$-$} \\[1mm]
        \midrule
        \multicolumn{4}{l}{(Results presented in \protect\citealt{adame2024desi3})} \\ \midrule
        BGS & {$-$} & {$0.983\pm0.019$} & {$-$} \\[1mm]
        LRG1  & {$-$} & {$0.979\pm0.011$} & {$0.915\pm0.037$} \\[1mm] 
        LRG2  & {$-$} & {$0.966\pm0.012$} & {$1.046\pm0.043$}\\[1mm]% 
        LRG3+ELG1 & {$-$} & {$0.998\pm0.008$} & {$1.026\pm0.028$} \\[1mm] %
        ELG2  & {$-$} & {$0.988\pm0.015$} & {$0.990\pm0.047$} \\[1mm] %
        QSO & {$-$} & {$1.002\pm0.026$} & {$-$} \\[1mm]
        \bottomrule
    \end{tabular}
\end{table}

We also show the correlation function data and best fit models for each tracer for our baseline fits in Figure~\ref{fig::results_YR1_datavsmodels_tracers_baseline}. Unlike the results presented in \cite{adame2024desi3}, the model here shows the fit to the data when $\beta_{\phi}$ is included in the analysis. For comparison, Figures 5 and 7 in \cite{adame2024desi3} show the model fits compared to the data for the standard BAO analysis. The $\chi^2$ goodness-of-fit and number of degrees of freedom of each tracer is given in Table~\ref{tab::results_year1chisquare} and in all cases these are comparable to those presented in Table 15 of \cite{adame2024desi3} and demonstrate excellent fits. 

\begin{table}
    \centering
    \caption{The $\chi^2$ and degrees-of-freedom (dof) for each of the Baseline fits to the different DESI tracers when $\beta_{\phi}$ is allowed to vary (corresponding to the data shown in Figure~\ref{fig::results_YR1_datavsmodels_tracers_baseline}). }
    \begin{tabular}{>{\raggedright\arraybackslash}p{2cm}c}
       \toprule
       \midrule
       \textbf{Tracer} & {$\chi^2/\mathrm{dof}$} \\ \midrule
       BGS & $15.8/14$  \\[2mm]
       LRG1 & $40.8/36$  \\[2mm]
       LRG2 & $40.6/36$ \\[2mm]
       LRG3+ELG1 & $31.6/36$ \\[2mm]
       ELG2 & $59.8/36$ \\[2mm]
       QSO & $28.8/14$ \\
       \bottomrule
    \end{tabular}
    \label{tab::results_year1chisquare}
\end{table}

\begin{figure*} 
    \begin{subfigure}[b]{0.485\textwidth} 
    \includegraphics[width=1\textwidth]{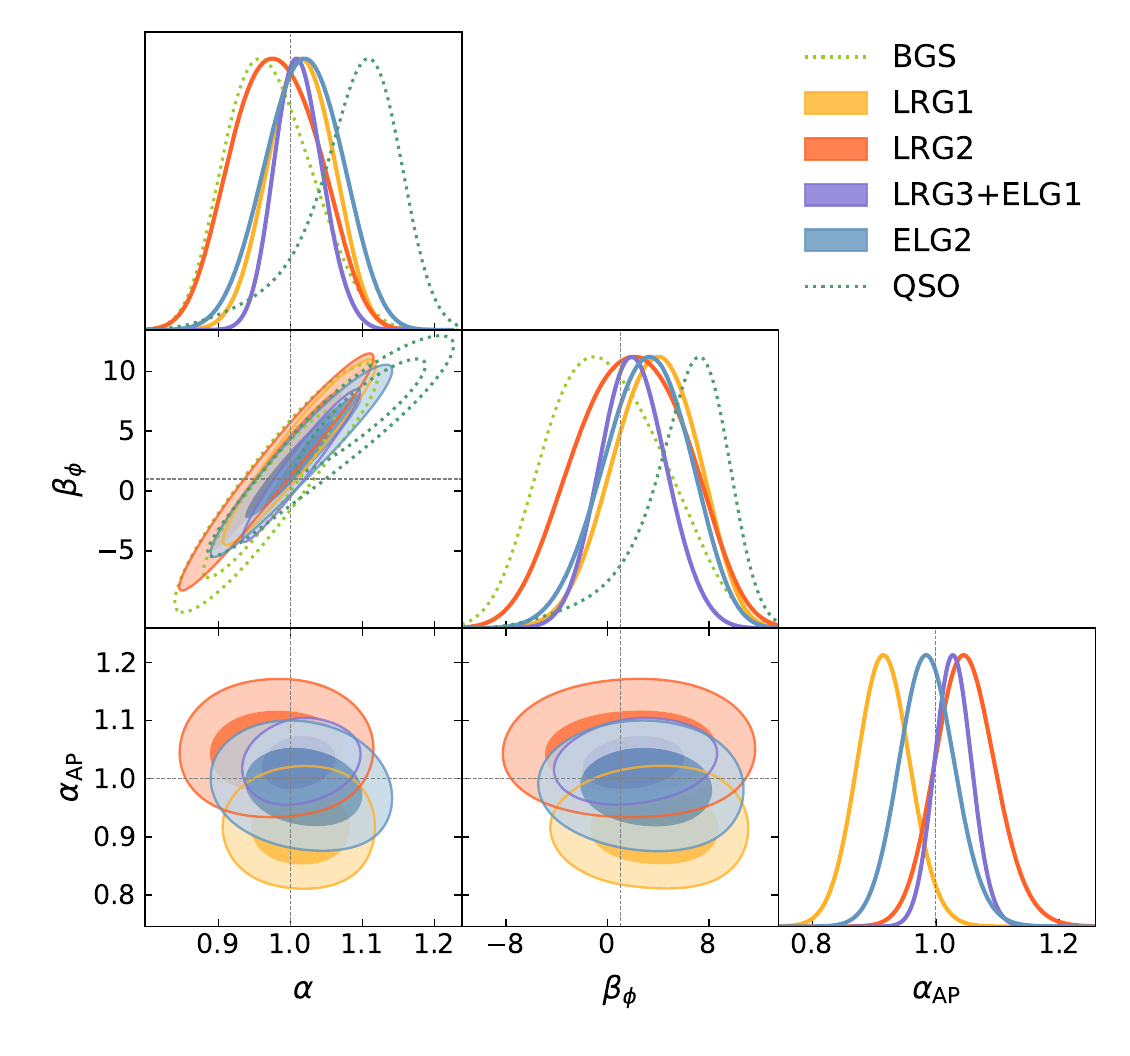}
    \end{subfigure} 
    \begin{subfigure}[b]{0.49\textwidth} 
    \includegraphics[width=1\textwidth]{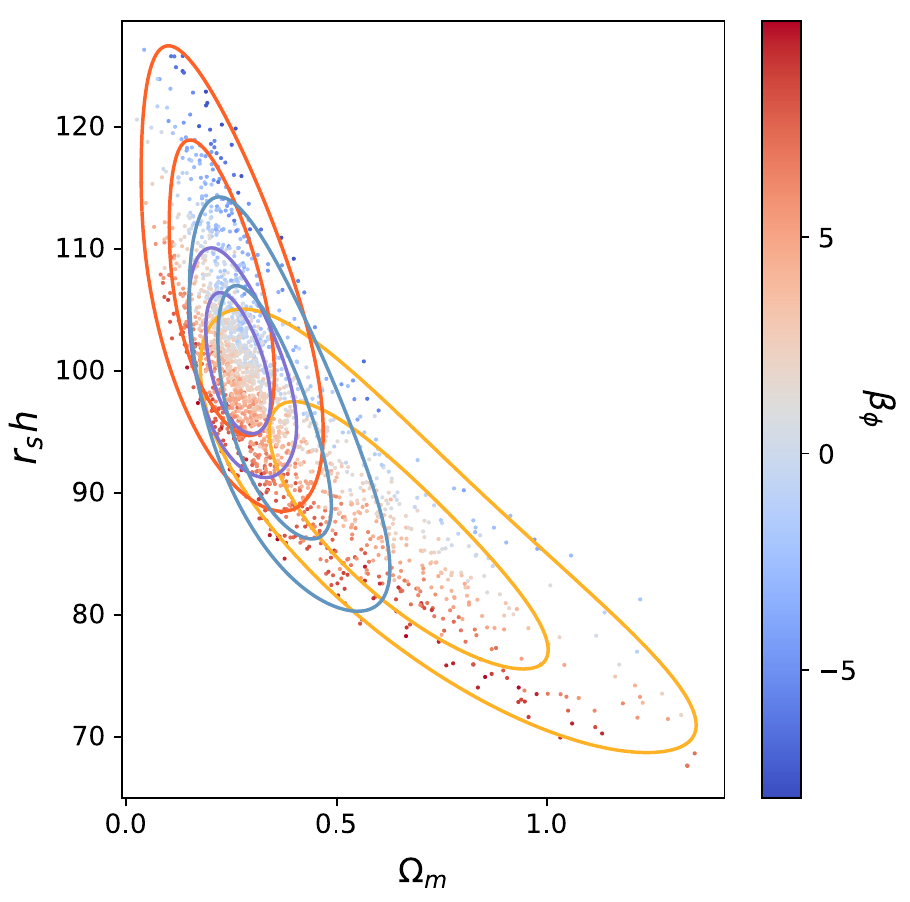}    
    \end{subfigure} 
    \caption{Left panel: contours for $\alpha$, $\alpha_{\mathrm{AP}}$ and $\beta_{\phi}$ for the baseline fits to the DESI DR1 data. The fits by QSOs and BGS which are not able to constrain $\beta_{\phi}$ well in the DR1 data are shown as dotted unfilled contours. Right panel: contours for $\Omega_{\rm{m}}$, $r_sh$ (Mpc) for the baseline fits to the DESI DR1 data, for individual tracers. The scatter points and colorbar show the value of $\beta_{\phi}$ which as a degeneracy with $r_sh$.} \label{fig::results_YR1_contours_alphabeta}
\end{figure*}

Figure~\ref{fig::results_YR1_contours_alphabeta} shows the individual fits to each tracer and the posterior distributions for our parameters of interest. The triangle plot on the left allows one to see the posterior distribution for each tracer for $\alpha$, $\alpha_{\rm{AP}}$ (excluding BGS and QSOs) and $\beta_{\phi}$. Each individual bin for the tracers gives relatively consistent fits on $\beta_{\phi}$ (we do not expect consistency for the $\alpha$, $\alpha_{\rm{AP}}$ as they are redshift dependent). The BGS and QSO constraints show the largest differences, but as mentioned previously, this is likely because QSOs and the BGS data individually are not able to constrain the $\beta_{\phi}$ parameter in addition to $\alpha$ strongly. Assuming a flat $\Lambda$CDM cosmological model, the anisotropic BAO constraints can also be uniquely mapped to a constraint on $\Omega_{m}$ and $r_sh$ (see equation~\ref{eq:alphaaptocosmo} and equation~\ref{eq:alphatocosmo}). These are shown in the right-hand panel of Figure~\ref{fig::results_YR1_contours_alphabeta}. As is shown in the DESI DR1 results in Figure 2 of \cite{adame2024desi}, the contours for $r_sh$ and $\Omega_{\rm{m}}$ have a rotating degeneracy direction for different tracers (due to the changing redshift of each tracer bin), but the constraints between different tracers are internally consistent and the combination allows for a tighter constraint on all parameters.

\subsection{Combined constraints and cosmological interpretation}
\begin{figure*}
    \begin{subfigure}[b]{0.44\textwidth} 
    \includegraphics[width=1\textwidth]{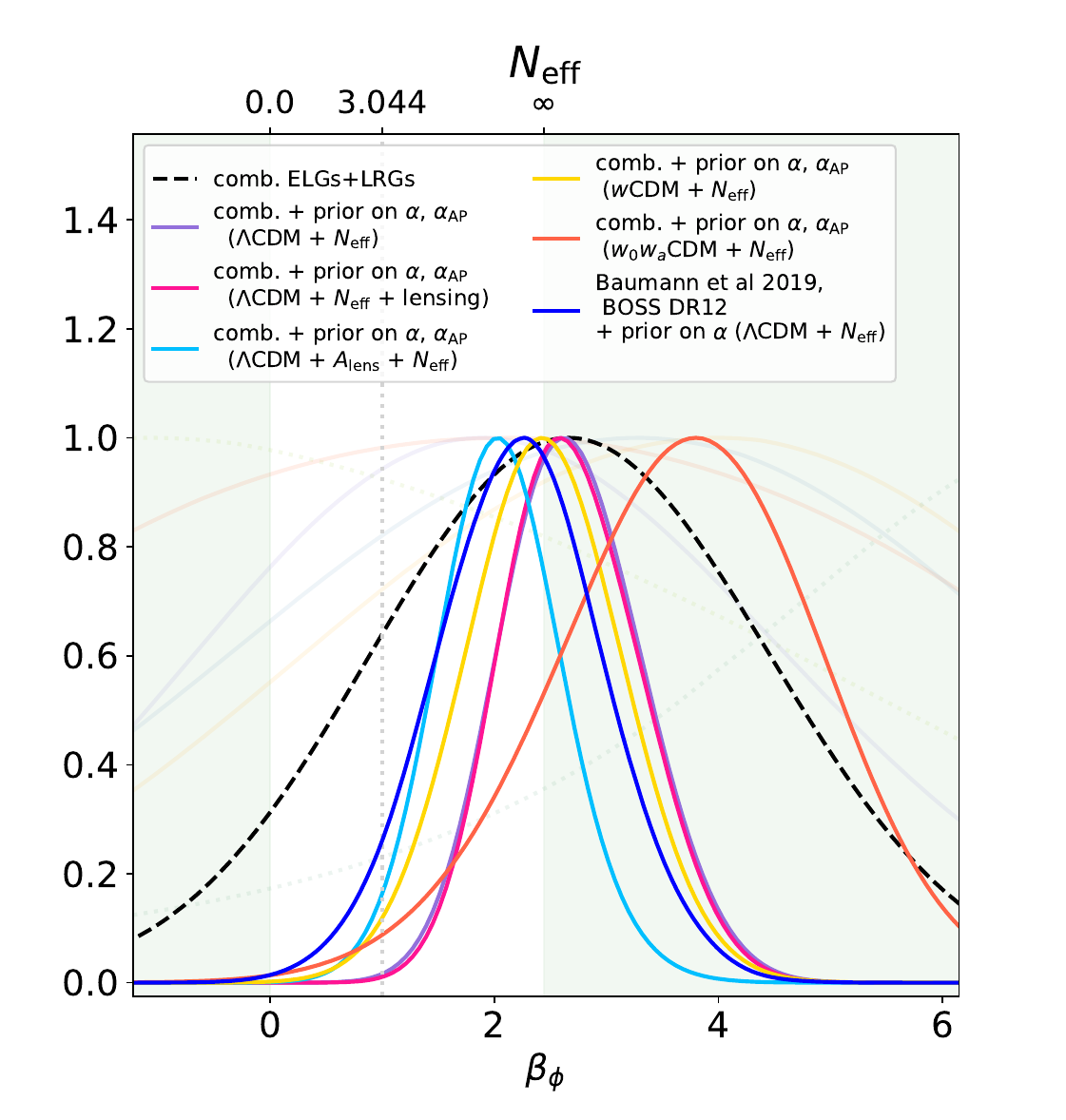}
    \end{subfigure} 
    \begin{subfigure}[b]{0.52\textwidth} 
    \includegraphics[width=1\textwidth]{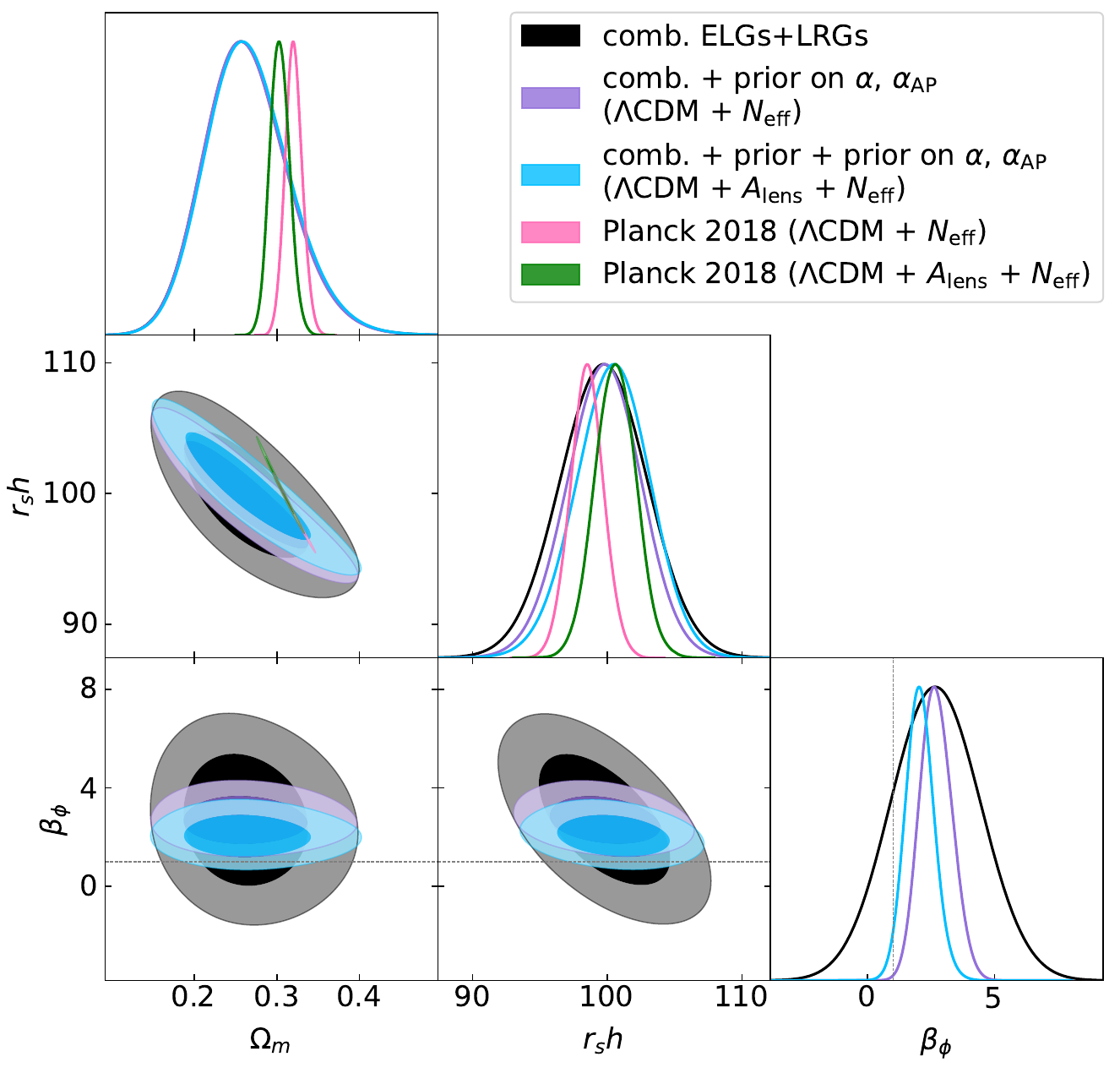}
    \end{subfigure}
    \caption{Left panel: the 1D posterior fits for $\beta_{\phi}$, including a combined fit to the tracers (excluding BGS and QSOs, black dashed line), and various combined fits when a CMB prior for $\alpha$ and $\alpha_{\rm{AP}}$ has been included by importance sampling (corresponding to the fits shown in Table~\ref{tab::results_year1summary}). The priors all use fits to the \textit{Planck} chains for TT+TE+EE + low$\ell$ + lowE. We include a prior from fits to $\Lambda$CDM + $N_{\mathrm{eff}}$ (purple, dashed line), the same but with lensing information included (pink, dashed), the case for a $\Lambda$CDM prior (blue), $\Lambda$CDM + $A_{\rm{lens}}$ (red) and finally $w$CDM (orange). The faint lines show the 1D posteriors of $\beta_{\phi}$ for each tracer that are shown in Figure~\ref{fig::results_YR1_contours_alphabeta}. The dark blue line shows the fit to BOSS DR12 by \citep{baumann2019first}. Right panel: a combined fit for $\Omega_{\rm{m}}$, $r_sh$ and $\beta_{\phi}$ to three LRG bins and the ELG2 bin is shown as a black contour. The fits by QSOs and BGS are not included here. We also show the fits when each tracer has a CMB prior for $\alpha$ and $\alpha_{\rm{AP}}$ by importance sampling using the \textit{Planck} chains for fits to TT+TE+EE + low$\ell$ + lowE in the context of $\Lambda$CDM + $N_{\mathrm{eff}}$. For comparison the \textit{Planck} chains are also shown for just $\Omega_{\rm{m}}$ and $r_sh$. } \label{fig::results_YR1_contours_cosmology}
\end{figure*}

As our different tracers give consistent cosmological results, we can combine them using the methodology described in section~\ref{sec::combinedfits} to combine the ELG and LRG mocks. The left panel of Figure~\ref{fig::results_YR1_contours_cosmology} focuses on just the 1-dimensional posterior for $\beta_{\phi}$ and includes the result for the combined fit to the tracers as a black-dashed line. The right panel of Figure~\ref{fig::results_YR1_contours_cosmology} shows the contours for the combined fits for $\Omega_{\rm{m}}$ and $r_sh$ when a flat $\Lambda$CDM model has been assumed. 

Additionally, we show various results when a prior on $\alpha$ and $\alpha_{\rm{AP}}$ from \textit{Planck} is included \citep[following the approach used by][]{baumann2019first} by importance sampling the fits to each tracer with the information from the \textit{Planck} 2018 MCMC chains for TT+TE+EE+ low$\ell$ + lowE.\footnote{This prior is \emph{not} the same as the prior included earlier from \textit{Planck} on $N_{\mathrm{eff}}$. This prior is only for the case we want to include information that impacts the constraints on $\alpha$ and $\alpha_{\rm{AP}}$, and allow the BAO data alone to constrain $\beta_{\phi}$. However we do only include information from \textit{Planck} on $\alpha$ and $\alpha_{\rm{AP}}$ from chains in with $N_{\mathrm{eff}}$ varies freely.} In these cases, the combined fit for $\beta_{\phi}$ is obtained using the same method as previously (equation~\ref{eq:combined}), but only after the individual posterior distribution function for each tracer bin has been importance sampled with the \textit{Planck} data. In other words, we have computed a new combined posterior as
\begin{align}
    P_{\mathrm{combined}}(\beta_{\phi}) & \propto \prod_{i} \int d\alpha_i\,d\alpha_{\mathrm{AP},i}\,P_{i}(\alpha_i, \alpha_{\mathrm{AP},i}, \beta_{\phi}|\xi_{i}) \nonumber \\
    & \qquad \qquad \times P_{\rm{Planck}}(\alpha_i, \alpha_{\mathrm{AP},i}|\Omega_{\rm{m}}, r_sh, w, N_{\rm{eff}}),
\end{align}
in which $\Omega_{\rm{m}}, r_sh$, $N_{\mathrm{eff}}$ and $w$ (the matter density, sound horizon, effective number of neutrino species and in some cases additional parameters such as the dark energy equation of state, EOS) are constrained by \textit{Planck} and $i$ is the set of DESI tracers. Generally for the $\Lambda$CDM case in which dark energy is a cosmological constant we assume the curvature density $\Omega_{\rm{k}} = 0$ and $\Omega_{\Lambda} = 1 - \Omega_{\rm{m}}$ where $\Omega_{\Lambda}$ drives the accelerated expansion of the Universe. Each density component has an EOS and this has $w$ fixed to $w = -1$ for a cosmological constant. When adding our \textit{Planck}-based priors we perform the same calculation as in equation~\ref{eq:alphaaptocosmo} and equation~\ref{eq:alphatocosmo}, but not necessarily limited to the $\Lambda$CDM case (in which case there is no longer a unique mapping from cosmological to BAO parameters). We can allow for a model in which the EOS of dark energy varies and thus $w$ is an additional free parameter. We also consider the case in which the dark energy EOS is parameterized by $w(a) = w_0 + w_a(1 - a)$ or when $A_{\rm{lens}}$ is a free parameter. This is discussed later in this section. 

By including the additional information from \textit{Planck} about $\Omega_{m}$ and $r_sh$ for the $\Lambda$CDM + $N_{\mathrm{eff}}$ case, the uncertainty on $\alpha$ and $\alpha_{\mathrm{AP}}$ is reduced for each tracer. Consequently, this allows for better constraints not only on $\alpha$ but on $\beta_{\phi}$ due to their strong degeneracy. This can be seen from the purple and pink dashed lines in the left-hand panel of Figure~\ref{fig::results_YR1_contours_cosmology}, where, interestingly, the best fit value for $\beta_{\phi}$ does not shift noticeably when the information from the \textit{Planck} $\Lambda$CDM + $N_{\mathrm{eff}}$ chains as prior information, with or without lensing information included. In including information from the \textit{Planck} chains, we do not fold in any explicit CMB information on $N_{\mathrm{eff}}$ in order to report only the constraining power available on $\beta_{\phi}$ from the physical shifting of the BAO wiggles; the constraints on $N_{\mathrm{eff}}$ from \textit{Planck} include additional information on $N_{\mathrm{eff}}$ from the CMB that is degenerate with other cosmological parameters, in the same way that there are additional signals in the galaxy clustering beyond the BAO phase shift that can constrain $N_{\mathrm{eff}}$ (see Figures~\ref{fig::effectalteringneff} and~\ref{fig::effectalteringneffcorr}).

\begin{table*}
    \centering
    \caption{\small Constraints on $\beta_{\phi}$ from the combined DESI DR1 BAO tracers, alongside cases when a prior on the BAO $\alpha$ parameters has been included from the \textit{Planck} data by weighting the chains from the BAO fits. We also show the number of (Gaussian) $\sigma$ away each best fit to $\beta_{\phi}$ is from zero ($N_{\mathrm{eff}}=0$) and unity ($N_{\mathrm{eff}}=3.044$).}
    \label{tab::results_year1summary}
    \begin{tabular}{lccccc}
        \toprule
        \midrule
        
        \textbf{Data} &  $\Omega_{m}$ & $r_sh$ (Mpc) & $\beta_{\phi}$ & {$\Delta \sigma (\beta_{\phi} > 0)$} & {$\Delta \sigma (\beta_{\phi} > 1)$} \\
        \midrule
        %\multicolumn{4}{l}{(DESI DR1 BAO)} \\ \midrule 
        DESI DR1 BAO & {$0.263^{+0.046}_{-0.054}$} & {$99.7\pm 3.2$} & {$2.70\pm1.70$} & {1.6} & {1.0} \\ [1mm]
        %\multicolumn{4}{l}{(BAO + \textit{Planck} prior)} \\ %\midrule 
        DESI DR1 BAO + \textit{Planck} prior $\alpha$,$\alpha_{\rm{AP}}$  & {$0.263^{+0.045}_{-0.054}$} & {$99.7\pm 2.7$} & {$2.70^{+0.60}_{-0.67}$} & {4.3} & {2.6} \\ [1mm]
    
        DESI DR1 BAO + \textit{Planck} prior $\alpha$,$\alpha_{\rm{AP}}$ w/ lensing & {$0.263^{+0.045}_{-0.054}$} & {$99.8\pm 2.7$} & {$2.69^{+0.59}_{-0.66}$} & {4.3} & {2.7} \\ [1mm]%\midrule

        DESI DR1 BAO + \textit{Planck} prior $\alpha$,$\alpha_{\rm{AP}}$ w/ varying $A_{\rm{lens}}$ & {$0.265^{+0.046}_{-0.054}$} & {$100.4\pm 2.7$} & {$2.05\pm 0.55$} & {3.7} & {1.9} \\ [1mm]%\midrule
        
        DESI DR1 BAO + \textit{Planck} prior $\alpha$,$\alpha_{\rm{AP}}$ w/ varying $w$ & {$-$} & {$-$} & {$2.44\pm0.70$} & {3.4} & {2.0} \\[1mm] 
        DESI DR1 BAO + \textit{Planck} prior $\alpha$,$\alpha_{\rm{AP}}$ w/ varying $w_0$, $w_a$ & {$-$} & {$-$} & {$3.7^{+1.2}_{-1.1}$} & {3.2} & {2.3} \\[1mm] 
       \bottomrule
    \end{tabular}
\end{table*}

Table~\ref{tab::results_year1summary} lists our combined constraints on $\beta_{\phi}$ from the DESI DR1 BAO data, and when the chains have been importance sampled with \textit{Planck} data. For the DESI data alone, the measurement of $\beta_{\phi}$ is consistent with unity within $1\sigma$, and suggests $\beta_{\phi} > 0$ at $> 68\%$ confidence. Including the \textit{Planck} prior on the BAO parameters (either with or without lensing) sharpens this constraint by a factor of $\sim 3$ while keeping the maximum posterior value fixed. This leads to a significant detection of a non-zero phase shift ($\beta_{\phi} > 0$ at $> 4\sigma$), but also leads to a constraint that prefers a value of $N_{\mathrm{eff}} > 3.044$ at greater than $95\%$ confidence. The maximum \textit{a posteriori} value is also large enough that it would correspond to an unphysical value of $N_{\mathrm{eff}}$, although we note that a wide range of $N_{\mathrm{eff}}$ is allowed given the non-linear mapping from this to $\beta_{\phi}$.

At first glance this then seems to suggest a phase shift in the BAO that, when combined with the \textit{Planck} prior, prefers either a larger number of effective relativistic species than predicted by the Standard Model, or the presence of other non-standard physics which introduces a phase shift in the BAO (e.g., non-standard neutrino physics, \citealt{kreisch2020neutrino}; or non-adiabatic primordial density fluctuations, \citealt{baumann2016phases}). A similar finding was reported by \cite{baumann2019first} for the BOSS DR12 BAO data, who found $\beta_{\phi} = 1.2 \pm 1.8$ from combining two redshift bins, and $\beta_{\phi}$ = $2.22 \pm 0.75$ after including a prior from \textit{Planck} by importance sampling. These two sets of results are remarkably consistent in central value and precision.

Compared to the fits from the BAO data alone or when priors from \textit{Planck} are included in the fits, \textit{Planck} alone prefers a larger $\Omega_{\rm{m}}$ and smaller $r_sh$. The discrepancy in the best fit for $\Omega_{\rm{m}}$ is slightly greater than the case for the fits to the BAO DR1 data fits when $\beta_{\phi}$ is fixed, in which case \cite{adame2024desi} reports $\Omega_{\rm{m}} = 0.295\pm0.015$. However, when $\beta_{\phi}$ is free the data prefers a value for $r_sh$ which is lower but more consistent with the best fit from \textit{Planck}. This may be related to the degeneracy that can seen between $r_sh$ and $\beta_{\phi}$ in Figure~\ref{fig::results_YR1_contours_cosmology}; this parameter degeneracy arises since both $r_sh$ and $\beta_{\phi}$ shift the BAO wiggles along the $k$-axis.

To investigate the impact of the \textit{Planck} priors on the fits to $\beta_{\phi}$ further, we also followed the tests done in \cite{baumann2019first} and included a \textit{Planck} prior for the $\Lambda$CDM model with $A_{\rm{lens}}$ (the lensing amplitude) allowed to be free in the analysis. We find a result $\beta_{\phi} = 2.05 \pm 0.55$, which is both a tighter constraint and closer to unity. The result is that our detection of a non-zero phase shift becomes now $>3\sigma$, but the tension with the Standard Model prediction is reduced somewhat from $2.6$ to $1.9\sigma$. This fit is still consistent with the result from the BOSS DR12 data at the $1\sigma$ level. \footnote{\citep{baumann2019first} find $\beta_{\phi} = 1.53 \pm 0.83$ for $\Lambda$CDM + $N_{\mathrm{eff}}$ + $A_{\rm{lens}}$.} During this analysis we did not have access to the \emph{Planck} 2021 (PR4) analysis \citep{tristram2024cosmological}, however we note in \cite{montefalcone2025free} who measure the phase shift in the CMB using both Planck 2018 and 2021 analyses find a shift that is more consistent with the standard model expectation.  

Lastly, while the \textit{Planck} data does tend to increase the tension between our $\beta_{\phi}$ and the Standard Model prediction, the prior from \textit{Planck} is based on chains from sampling in the $\Lambda$CDM model, while DESI data (in combination with CMB and Type Ia Supernovae) moderately prefer a $w_0w_a$CDM (time-varying Dark Energy) model \citep{adame2024desi}. Given that we are including a prior from a model that is not preferred by DESI constraints on the BAO scaling parameters, this might encourage the combined fit to $\beta_{\phi}$ from DESI with a $\Lambda$CDM \textit{Planck} prior to take a higher than expected value of $\beta_{\phi}$. As such we include fits from a full MCMC fit to a time-varying Dark Energy model with varying $N_{\mathrm{eff}}$. We present results for the $w$CDM model (in which case we have the normal flat $\Lambda$CDM model + $N_{\mathrm{eff}}$ but $w$ is allowed to vary as a single free parameter). We find this tends to allow the best fit value of $\beta_{\phi}$ to shift to a value that also reduces the tension with the standard model with $\beta_{\phi} = 2.44\pm0.70$. This may be due to the additional flexibility in the model. In the case for the time-varying EOS model $w_0w_a$CDM + $N_{\mathrm{eff}}$, where the additional parameters $w_0$ and $w_a$ vary in the EOS $w(a) = w_0 + (1-a)w_a$, we find that the additional free parameters lead to weaker constraints and has slightly greater tension with the theoretical expectation for $N_{\mathrm{eff}}$, with $\beta_{\phi} = 3.7^{+1.2}_{-1.1}$.

\section{Conclusion}\label{sec::conclusions}

In this work, we have applied the scheme derived in \cite{baumann2019first} to measure the phase shift in the BAOs. This phase shift is induced by free-streaming particles that propagate faster than the sound speed of the primordial plasma and thus allows one to measure the impact of free-streaming neutrinos that contribute to $N_{\rm{eff}}$ in the early Universe. 

Compared to the analysis on BAOs by \cite{baumann2019first}, we have extended the phase shift scheme for an anisotropic BAO fitting analysis, which we have applied to the DESI DR1 BAO data for galaxies and quasars. In the process, we have also extended the publicly available codes \textit{desilike} to create the fits and likewise \textit{Barry} to validate the fitting methodologies used between two separate codes. We tested our pipeline with \textit{desilike} and \textit{Barry} on highly precise cubic box simulations to ensure the robustness of the fitting schemes when applied to data with low statistical noise. Further to this we tested the pipeline and codes on realistic mocks for the various tracers in the DESI DR1 data release before applying the code to the real DR1 data. 

Our analysis on the highly precise cubic box simulations has highlighted the need to address systematics in the BAO fitting methodology for future constraints with the 5 year DESI data. Further investigation will be required in order to determine a more robust way to measure the phase shift for different broadband fitting methodologies for the power spectrum. We also expect there may be a need to address other differences seen in the fits when varying the analysis pipeline if the measurements of the BAO distortion parameters and $\beta_{\phi}$ are larger relative to a reduced statistical uncertainty in future measurements. Potentially, it will also be necessary to consider in greater detail the impact of the choice of fiducial cosmology or template cosmology. Further to this, while allowing $\beta_{\phi}$ to vary significantly weakens constraints on $\alpha$ due to the strong degeneracy between these parameters, it may be interesting to consider where fixing $\beta_{\phi}$ could lead to small biases in $\alpha$, particularly as we continue to report more precise measurements with upcoming data from DESI.

We have measured the phase shift in the DR1 DESI data from the combination of LRGs and ELGs across various redshift bins to be $\beta_{\phi} = 2.7 \pm 1.7 $. This is larger than the measurement of $\beta_{\phi} = 1.2 \pm 1.8$ in \cite{baumann2019first}, however both measurements agree within the $1\sigma$ uncertainties. When we add a prior on $\alpha$ and $\alpha_{\rm{AP}}$ from \textit{Planck} 2018 measurements of the CMB, we obtain $\beta_{\phi} = 2.7^{+0.60}_{-0.67}$, which as before is larger than the value of $2.22 \pm 0.75$ given in \cite{baumann2019first} with the same prior included. In cosmological models with only Standard Model particles the phase shift can be interpreted as due to $N_{\rm{eff}}$. However here the phase shift measurements of $\beta_{\phi} \gtrsim 2.44$ do not correspond to a physical value for $N_{\rm{eff}}$. The amplitude of the phase shift may hint at physics beyond the standard model since it is not entirely consistent with an interpretaton via $N_{\mathrm{eff}}$. However, we emphasize here that the measurement of $\beta_{\phi}$ is for a physical phase shift that is present in the BAOs, and in a non-standard model could be impacted by non-standard neutrino physics \citep{kreisch2020neutrino} or by non-adiabatic fluctuations \citep{baumann2016phases}. Furthermore, our measurements of $\beta_{\phi}$ from DESI BAOs alone are still consistent with physical values of $N_{\rm{eff}}$ at the $1\sigma$ level. 

In the future, the analysis can potentially be applied to the DESI Y3 or DESI Y5 data. \cite{baumann2019first} forecasts that with the DESI Y5 data it may be potential to obtain $\sigma(\beta_{\phi}) \sim 0.3$, which is a factor of $\sim 6$ better than the constraints reported here for the BAO only data, although these forecasts may not include the impact of allowing $\alpha_{\rm{AP}}$ to vary freely as is done in this work. In order to provide competitive constraints on future data with DESI Y5, it will be necessary to begin to improve the systematics on our measurements that were analysed in detail in section~\ref{sec::validationmocks}. 

In future work, it will also be interesting to consider the impact of a phase shift due to particles that may have different properties to Standard Model neutrinos, such as interacting neutrinos. It may be expected that such neutrinos could change the functional form of the phase shift that is given by equation~\ref{eq::phaseshift_kdependence} \citep[which has been studied in][]{choi2018probing} due to the fact that neutrino interactions can change the epoch at which they decouple and free-stream \citep{kreisch2020neutrino, camarena2023confronting}. This may be particularly interesting in the future when stronger constraints on $\beta_{\phi}$ are possible, particularly in light of results that find a bimodal distribution for a parameter that characterises the interaction strength of strongly interacting neutrinos \citep{camarena2023confronting, kreisch2024atacama}. It may also be interesting for future work to consider the ability of DESI+\textit{Planck} to constrain the phase shift in the context of more models beyond $\Lambda$CDM, or consider adding information from CMB probes for the phase shift induced by free-streaming relics, such as the constraints found by \cite{follin2015first}.

\section*{Acknowledgements}\label{sec::acknowledges}

This research has made use of NASA's Astrophysics Data System bibliographic services, the astro-ph pre-print archive at \url{https://arxiv.org/} and the python libraries \texttt{MATPLOTLIB}, \texttt{GETDIST}, \texttt{CHAINCONSUMER}, \texttt{NUMPY}, \texttt{SCIPY} and \texttt{PANDAS} \citep{hunter2007matplotlib, Lewis2019xzd, Hinton2016, harris2020array, 2020SciPy-NMeth, mckinney-proc-scipy-2010}. 

This material is based upon work supported by the U.S. Department of Energy (DOE), Office of Science, Office of High-Energy Physics, under Contract No. DE–AC02–05CH11231, and by the National Energy Research Scientific Computing Center, a DOE Office of Science User Facility under the same contract. Additional support for DESI was provided by the U.S. National Science Foundation (NSF), Division of Astronomical Sciences under Contract No. AST-0950945 to the NSF’s National Optical-Infrared Astronomy Research Laboratory; the Science and Technology Facilities Council of the United Kingdom; the Gordon and Betty Moore Foundation; the Heising-Simons Foundation; the French Alternative Energies and Atomic Energy Commission (CEA); the National Council of Humanities, Science and Technology of Mexico (CONAHCYT); the Ministry of Science, Innovation and Universities of Spain (MICIU/AEI/10.13039/501100011033), and by the DESI Member Institutions: \url{https://www.desi.lbl.gov/collaborating-institutions}. Any opinions, findings, and conclusions or recommendations expressed in this material are those of the author(s) and do not necessarily reflect the views of the U. S. National Science Foundation, the U. S. Department of Energy, or any of the listed funding agencies.

The authors are honored to be permitted to conduct scientific research on Iolkam Du’ag (Kitt Peak), a mountain with particular significance to the Tohono O’odham Nation.

This work used the DiRAC@Durham facility managed by the Institute for Computational Cosmology on behalf of the STFC DiRAC HPC Facility (www.dirac.ac.uk). The equipment was funded by BEIS capital funding via STFC capital grants ST/K00042X/1, ST/P002293/1, ST/R002371/1 and ST/S002502/1, Durham University and STFC operations grant ST/R000832/1. DiRAC is part of the National e-Infrastructure.

The authors thank Willem Elbers for providing full MCMC fits to the \textit{Planck} 2018 data for various cosmological models. The authors also thank Willem Elbers, Ariel Sanchez and Seshadri Nadathur for their substantial feedback that has helped to improve the writing of this paper. AW thanks David Camarena for interesting discussion, facilitated by the Galileo Galilei Institute of Theoretical Physics in Florence, Italy at the \emph{Neutrino Frontiers} workshop in 2024. AW also thanks Benjamin Wallisch for useful suggestions. AW, TD and CH acknowledge support from the Australian Government through the Australian Research Council’s Laureate Fellowship funding scheme (FL180100168). 
HR, MV, and SF are supported by PAPIIT IN108321, PAPIIT IN116024, and PAPIIT IN115424. 

%%%%%%%%%%%%%%%%%%%%%%%%%%%%%%%%%%%%%%%%%%%%%%%%%%
\section*{Data Availability}
The data to produce the figures in this work can be found at \url{https://zenodo.org/records/14311742}. The code used to produce fits with \emph{Barry}, including modifications for the phase shift amplitude, is available at \url{https://github.com/abbew25/Barry}. The \emph{desilike} code to produce fits with the phase shift amplitude can be accessed at \url{https://github.com/cosmodesi/desilike}.



\bibliographystyle{mnras}
\bibliography{bibliography} % if your bibtex file is called example.bib

%\begin{thebibliography}{99}
%\end{thebibliography}

%%%%%%%%%%%%%%%%%%%%%%%%%%%%%%%%%%%%%%%%%%%%%%%%%%

%%%%%%%%%%%%%%%%% APPENDICES %%%%%%%%%%%%%%%%%%%%%

\appendix

\section{Validating the anisotropic pipeline by comparison to Baumann et al (2019)}\label{app::fitsbaumannvalidation}

\begin{figure}
    \centering
    \begin{subfigure}[b]{0.47\textwidth}
    \centering
    \includegraphics[width=1.0\textwidth]{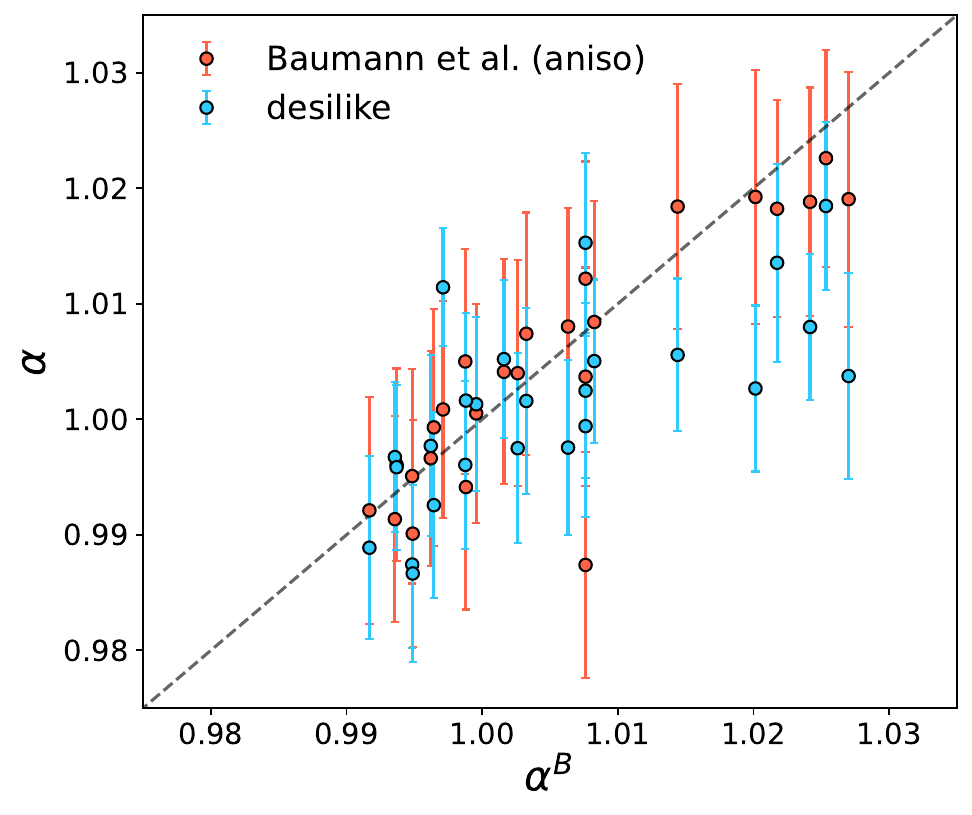}
    \end{subfigure}
    \begin{subfigure}[b]{0.475\textwidth}
    \centering
    \includegraphics[width=1.0\textwidth]{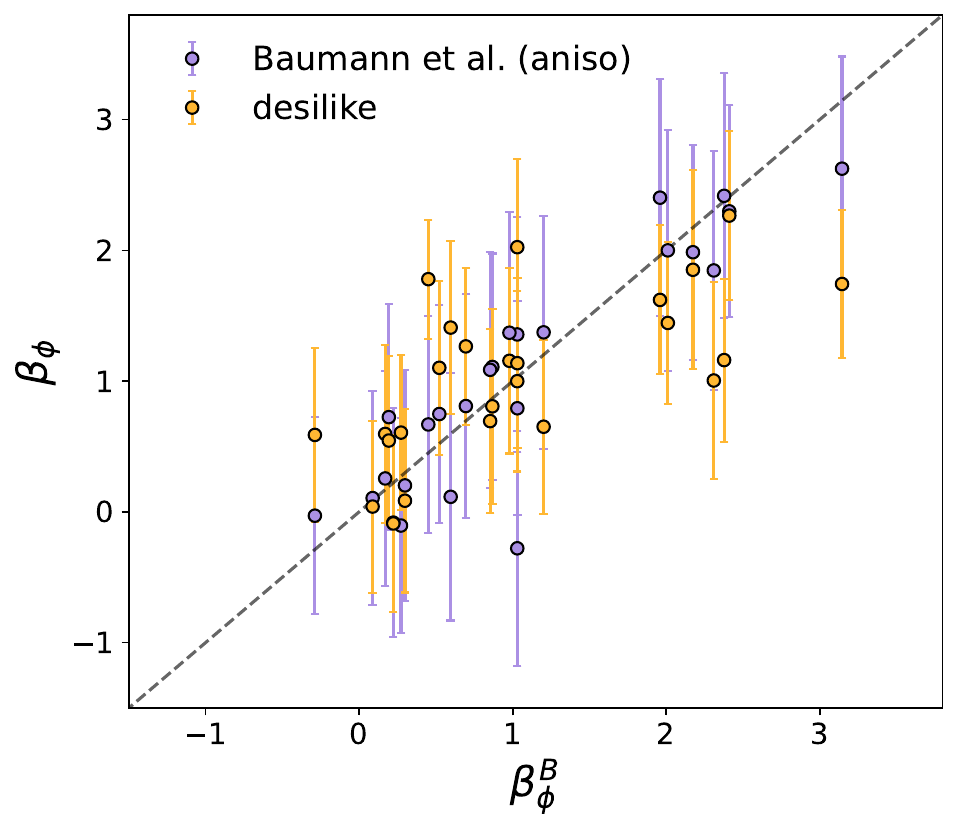}
    \end{subfigure}
    \caption{\small Results of fitting $\alpha$ (top panel) and $\beta_{\phi}$ (lower panel) for 25 mock realisations using the Pipeline of \protect\cite{baumann2019first}, compared to an extended version of the Baumman pipeline for anisotropic fitting, and additionally comparing to the \textit{desilike} pipeline. }
\label{fig::baummannpipelinefits_vs_desilike}
\end{figure}

We show the fits produced by following the isotropic BAO pipeline by \cite{baumann2019first} with $\beta_{\phi}$ allowed to vary. Fits were produced in configuration space, and compared to the results from the \emph{desilike} pipeline (following an anisotropic fitting scheme) and an extended version of the approach of \cite{baumann2019first} for an anisotropic fitting scheme. The pipeline of \cite{baumann2019first} in configuration space uses a minimizer to find the best fit. The errors on the fits are evaluated by calculating the $\chi^2$ on a grid in $\{\alpha, \beta_\phi \}$ parameters to map out the likelihood surface. The anisotropic extension of this scheme calculates the best fit in a similar way but extends the parameter space to $\{\alpha, \beta_\phi, \alpha_{\rm{AP}} \}$. The approach by \cite{baumann2019first} uses the nonlinear model for the correlation function described in \cite{vargas2018}. However in their work, the Fourier-space fits use the same modelling as this work, following the methodology of \cite{beutler2017clustering}. 
The results of the fits from each code are compared in Figure~\ref{fig::baummannpipelinefits_vs_desilike}.

\section{Additional comparisons from \textit{Barry}}\label{app::fitsmocksbarrybetafirstgen_spline_pkvsxi}

\begin{figure}
    \centering
    \begin{subfigure}[b]{0.48\textwidth} 
    \includegraphics[width=1\textwidth]{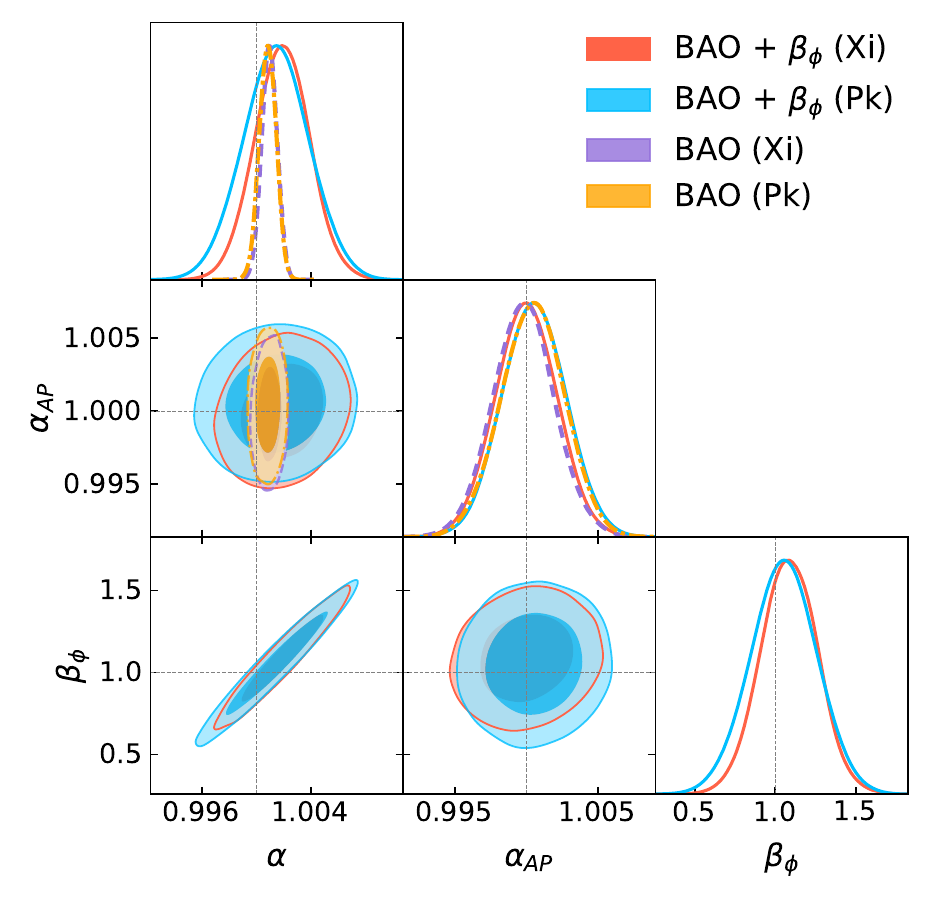}
    \caption{Our fits with spline broadband method.}
    \end{subfigure} 
    \begin{subfigure}[b]{0.48\textwidth} 
    \includegraphics[width=1\textwidth]{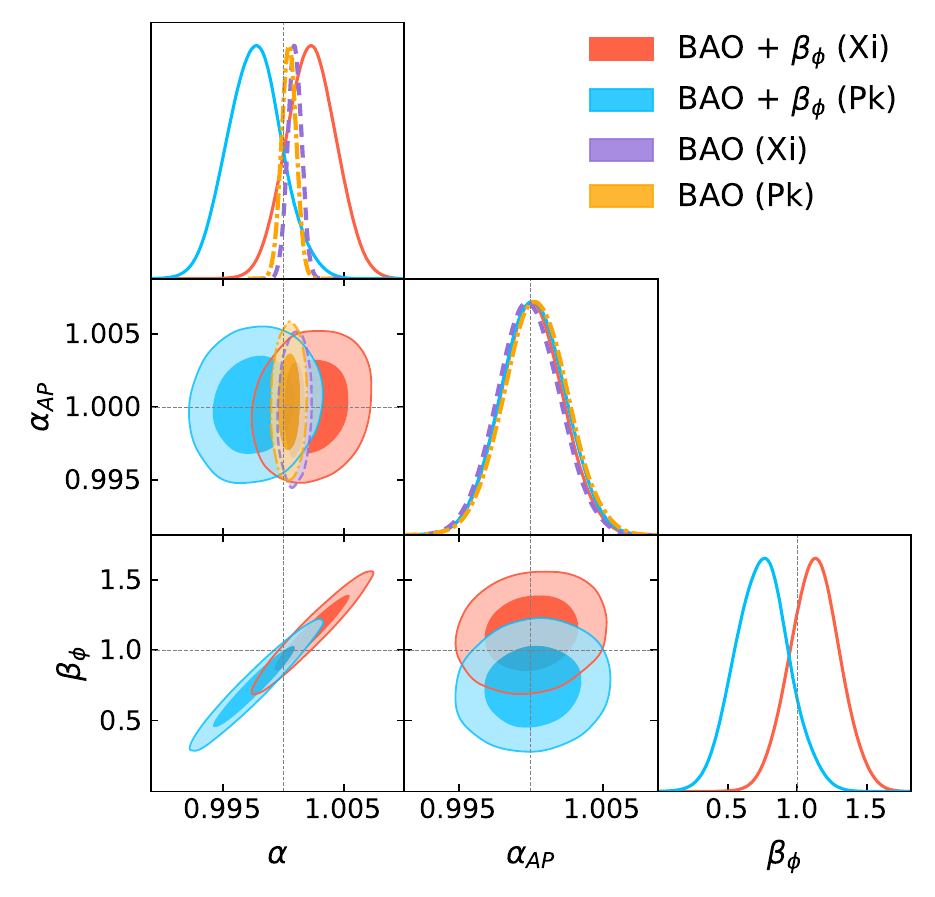}
    \caption{Our fits with polynomial broadband method.}
    \end{subfigure} 
    \caption{Comparison of the power spectrum and correlation function, for standard BAO fits with CV ELGs (first-generation snapshots at $z=1.1$) and for fits when the phase shift is varied. Here however, we show the difference when the spline broadband methodology is used.}
    \label{fig::comparison_XiPk_mocks_splinepoly} 
\end{figure}

In Figure~\ref{fig::comparison_XiPk_mocks_splinepoly} we show an additional fit for the mock mean of the first-generation control-variate mocks for ELGs with $z=1.1$, to compare the power spectrum and correlation function fits for the two broadband methodologies, and also to compare the results to the standard BAO fitting procedure. 

% \section{Combined fits to the mocks}\label{app::fitsmocksbarrydesilikecombined}

% Figure~\ref{fig::combined_fits_on_mocks} the combined fits for $\beta_{\phi}$ for the second-generation mocks for LRGs and ELGs (all bins except QSOs and BGS) with \textit{desilike} and \emph{Barry}. The fits here are done for correlation function mocks post-recon. The average absolute difference of each mock fit to the combination of the tracers, for fits to the tracers done using \textit{desilike} and \textit{Barry} is $\Delta \beta_{\phi} = 0.713$ (or $\frac{\Delta}{\sigma} = 0.485$ where $\sigma$ is the weighted standard error of the mean of the combined fits). The absolute difference
% between the fits to the mock mean data (average correlation function of 25 mocks) is $\Delta \beta_{\phi} = 0.38$. 
% \begin{figure}
%     \centering
%     \includegraphics[width=0.4\textwidth]{figures_desilike_barry_secondgen/combined_fits_barry_desilike.pdf}
%     \caption{Fits to $\beta_{\phi}$ from the combinations of the LRGs and ELGs (all tracer bins except QSOs and BGS) with the second-generation mocks, for \textit{desilike} (black points) and \textit{Barry} (white points). The shaded regions show the (weighted) 1-standard deviation about the weighted mean for the fits to individual mocks, the fit to the mean of the mocks is shown at the bottom of the plot.}
%     \label{fig::combined_fits_on_mocks}
% \end{figure}
%%%%%%%%%%%%%%%%%%%%%%%%%%%%%%%%%%%%%%%%%%%%%%%%%%

% Don't change these lines
\bsp	% typesetting comment
\label{lastpage}
\end{document}